\documentclass[11pt]{article}
\usepackage{amsmath,amssymb,jheppub,graphicx}

    \usepackage{etoolbox}
    \makeatletter
    \patchcmd{\maketitle}{\@fpheader}{}
    \makeatother
    
\vspace*{0.0 mm}

\usepackage{verbatim}
\usepackage{bm} 
\usepackage{slashed}
\usepackage{mathdots}
\usepackage{color}%
\def\hhref#1{\href{http://arxiv.org/abs/#1}{arXiv:#1}} 

\allowdisplaybreaks[1]

\usepackage[all,cmtip]{xy}

\def\Z{\mathbb{Z}} 
\def\R{\mathbb{R}} 
\def\C{\mathbb{C}} 
\def\P{\mathbb{P}} 
\def\K{{\cal{K}}}

\def\be{\begin{equation}}
\def\ee{\end{equation}}

\def\Dslash{{\rlap{\raise 1pt \hbox{$\>/$}}D}}

\def\Z{\mathbb{Z}} 
\def\R{\mathbb{R}} 
\def\C{\mathbb{C}}

\def\Im{\text{Im}}

\def\tr{\text{tr}}

\def\v{{\vee}}

\def\hat{\widehat}
\def\bar{\overline}

\def\a{{\alpha}}
\def\ba{{\bar\a}}

\def\G{{\Gamma}}
\def\d{{\delta}}

\def\coeff#1#2{{\textstyle {\frac {#1}{#2}}}}

\def\half{\coeff 12}

\title{
Resurgence and Dynamics of $\bm {O(N)}$ and Grassmannian Sigma Models
 } 
\author[1]{Gerald~V.~Dunne}
\author[2]{and Mithat \"Unsal}
\affiliation[1]{Department of Physics, University of Connecticut, Storrs CT 06269-3046, USA}
\affiliation[2]{Department of Physics, North Carolina State University, NC 27695-8202, USA}
\emailAdd{gerald.dunne@uconn.edu}
\emailAdd{unsal.mithat@gmail.com}
\abstract{   We study the non-perturbative dynamics of the two dimensional  ${O(N)}$  and Grassmannian sigma models by using compactification with twisted boundary conditions on $\mathbb R \times S^1$, semi-classical techniques  and resurgence. 
While the $O(N)$ model has no instantons for $N>3$,   it has (non-instanton) saddles on $\R^2$, which we call 2d-saddles.  On  $\mathbb R \times S^1$, 
the resurgent relation between perturbation theory and non-perturbative physics is encoded in  new saddles, which are associated with the affine root system of the ${\frak o}(N) $ algebra. These events may be viewed as  fractionalizations of the 2d-saddles.   
The first beta function coefficient, given by the dual Coxeter number, can then be intepreted as the sum of the  multiplicities (dual  Kac labels)  of these fractionalized objects. 
Surprisingly, the new saddles in $O(N)$ models in compactified space are in one-to-one correspondence with monopole-instanton saddles in $SO(N)$ gauge theory on 
$\R^3 \times S^1$.  
The Grassmannian sigma models  ${ \rm Gr}(N, M)$ have 2d instantons, which fractionalize into $N$ kink-instantons.  The small circle dynamics of both sigma models can be described as a dilute gas of the one-events and two-events, bions.  One-events are the leading source of a variety of non-perturbative effects, and produce the strong scale of the 2d theory in the compactified theory.  We show that in both types of sigma models the neutral bion emulates the role of  IR-renormalons. We also study the topological theta angle dependence in both the $O(3)$ model and  ${ \rm Gr}(N, M)$, and describe the  multi-branched structure of the observables  in terms of the theta-angle dependence of the saddle amplitudes. 
}

\begin{document}
\maketitle

\vfill

\eject

\allowdisplaybreaks
\section{Introduction}

There is now growing evidence that resurgent trans-series expansions may provide a non-perturbative continuum definition of quantum field theory (QFT), at least in their semi-classical regimes  \cite{Dunne:2012ae,Cherman:2013yfa,Argyres:2012ka,Misumi:2014rsa},  
 and possibly in the strong coupling regime where the operator product expansion is interpreted as a trans-series. Resurgent trans-series encode infinite families of relations between distinct sectors, perturbative and non-perturbative. For example, in prototypical quantum mechanical systems like the double-well  and  periodic Mathieu potentials, well-studied models for instantons and non-perturbative physics, one finds that in fact {\it all} non-perturbative information, to all orders, is encoded in a subtle way in perturbation theory \cite{Dunne:2013ada}.  (For a recent direct confirmation, see \cite{Escobar-Ruiz:2015nsa}.)
  
 Resurgence has found many applications, for example in differential equations, dynamical systems, and fluid mechanics, and it is now widely regarded as a universal approach to asymptotic problems with a large or small parameter \cite{Costin:2009,delabaere, Sauzin}. In addition to the QFT applications  \cite{Dunne:2012ae,Cherman:2013yfa,Misumi:2014rsa,Argyres:2012ka}, resurgent analysis has recently been applied to a variety of problems, such as matrix models, Chern-Simons theories and topological strings \cite{Marino:2007te,Pasquetti:2009jg,Aniceto:2011nu,Marino:2012zq},  ABJM theory \cite{Kallen:2013qla},  the holomorphic anomaly \cite{Santamaria:2013rua}, supersymmetric localization \cite{Aniceto:2014hoa}, fractionalized classical solutions \cite{Dabrowski:2013kba,Bolognesi:2013tya,Bruckmann:2014sla,Misumi:2014bsa,Nitta:2015tua,Nitta:2}, Lefschetz thimbles \cite{Tanizaki:2014tua},  Nekrasov partition functions \cite{Basar:2015xna} and hydrodynamics \cite{Heller:2015dha}.

Here we apply the resurgence formalism and the physical principle of adiabatic continuity on $\R \times S^1$ to the simplest  asymptotically free QFT with a mass gap, the $O(N)$ non-linear sigma model in two spacetime dimensions (2d).   The $O(N)$ model is of particular interest because while  it is soluble at large $N$, and  is  integrable  at finite-$N$
   \cite{Zamolodchikov:1978xm,Novikov:1984rf,Polyakov:1983tt,Hasenfratz:1990zz,Beneke:1998eq,abdalla, Volin:2009wr}, the   non-perturbative structure is  not yet fully understood. 
   In particular,  integrability {\it assumes} the    existence of a mass gap, which needs to be shown at finite-$N$.
   Furthermore, the model has no instantons for $N\geq 4$ in 2d, unlike   4d  gauge theories.   In standard textbooks, this is often presented as a point where good analogy between 4d gauge theory and 2d $O(N)$ models  break down, deeming  them useless  as a  2d laboratory to understand non-perturbative effects.    (As shown in this paper, this point of view 
   turns out to be superficial.) 
  It is actually  expected that the $O(N)$ model may possess  a different non-perturbative structure compared to the 2d $\mathbb C\mathbb P^{N-1}$ models, and 4d gauge theories, which do have instantons.       We hope to provide new insights into these problems. 
  We also generalize  our earlier work on two-dimensional  
  $\mathbb C\mathbb P^{N-1}$ models to the Grassmannian models ${ \rm Gr}(N, M)$. The ${ \rm Gr}(N, M)$ models are interesting because the leading beta function coefficient is $\beta_0=N$, independent of $M$, and  we show that the non-perturbative physics follows very closely that of  $\mathbb C\mathbb P^{N-1}\equiv { \rm Gr}(N, 1)$, studied in \cite{Dunne:2012ae}.

Non-perturbative semi-classical physics in gauge theories and sigma models is commonly identified
with homotopy arguments, such as the following:
the  $O(3)$ model in 2d
has instantons because $\pi_2(S^2)  = \Z$.
But $\pi_2(S^{N-1}) = 0$ for $N \geq 4$, so the $O(N)$ model with $N \geq 4$ has no instantons.
Contrast this with  $\C \P^{N-1}$ models, for which  $\C \P^{1} \equiv O(3)$, but
 $\C \P^{N-1}$ has instantons for all $N\geq 2$ \cite{zak-book,abdalla}.      
 However, the large-$N$ solution of the $O(N)$  model  shows that the Borel plane structure of perturbation theory associated with  $O(N)$  is identical to  that of the $\C \P^{N-1}$  model  \cite{David}. Thus, at least in  the large-$N$ limit, perturbation theory must have identical structure in these two theories, reminiscent of the large-$N$  orbifold-orientifold equivalences in gauge theories \cite{Armoni:2003gp,Kovtun:2004bz}. According to resurgence theory, the identical structure of perturbation theory must be mirrored in the non-perturbative saddles in the problem.  We show that the key to understanding this apparent puzzle is that, although the $O(N)$ model with $N \geq 4$ has no instantons, it has smooth finite action classical solutions of the second-order classical equations of motion [they are saddle points, not minima, of the action], and these play an important role in non-perturbative physics.
 
 Our motivation is based on our recent work in Yang-Mills and $\C \P^{N-1}$ \cite{Argyres:2012ka,Dunne:2012ae}, where spatial compactifications on $\R^d \times S^1$ with appropriate twisted boundary conditions brought these asymptotically free theories into a calculable semi-classical regime which is  connected adiabatically to the infinite volume limit
 in the sense of global symmetries, and universality. 
  In particular, for  $\C \P^{N-1}$ it was shown explicitly that the leading ambiguity in the Borel plane of perturbation theory [in the compactified theory] corresponds to neutral bions, the semi-classical realization of the IR-renormalons,  and is cancelled by an ambiguity in non-perturbative bion amplitudes. This provides  an explicit realization of resurgence in a non-trivial QFT.   This analysis was based on bions, correlated fractional  instanton/anti-instanton molecules with action $\frac{2S_I}{N}$, where $S_I$ is the action of the instanton on $\R^2$. 
  Here we address the important question: what happens in a theory without instantons?  In fact, this  question was addressed in earlier work on the $SU(N)$ principal chiral model, which also has no instantons, and where the 2d uniton saddle (a non-BPS solution to second order equations of motion)  fractionates into $N$ fracton constituents \cite{Cherman:2013yfa, Nitta:2}. However, this puzzle has remained unanswered for the $O(N \geq 4)$ vector model in 2d, the simplest 
 asymptotically free QFT.\footnote{Ref.~\cite{Nitta:2015tua}  considers generalization of $O(3)$ model on $\R^2$ to $O(N)$ model in $\R^{N-1}$, 
 so that instantons exists at any $N$ because  $\pi_{N-1}(S^{N-1})  = \Z$.    Ref.~\cite{Nitta:2015tua}  finds the  instantons as well as fractionalized instantons for general $N$. Here,  we address only  $O(N \geq 4)$ models  with no instantons, on two-manifolds $\R^2$ and  $\R \times S^1$. }   
There are many related questions:
(i) Are there any non-perturbative saddles in $O(N\geq 4)$ model? 
(ii) If there are, what are the actions associated with these saddles?  Are they related to the 2d strong (mass gap) scale   $ \Lambda$ 
 where  $ \Lambda= \mu e^{-\frac{2 \pi}{g^2(\mu) \beta_0}}$, 
 and renormalon singularities? (iii) Are there similarities between $O(N)$ gauge theory and $O(N)$  non-linear sigma models? In this paper, we address all these questions.

\subsection{Results for  $O(N)$-model} 
  In the $O(N)$ sigma model on $\R\times S^1$ endowed with twisted boundary conditions, the leading non-perturbative saddles are   fractional kink-instanton events ${\cal K}_j $ associated with the affine root system of  the $\frak  o(N)$-algebra.  At second order, there are charged and neutral bions,  ${\cal B}_{ij} = [  {\cal K}_i \overline {\cal K}_j   ] $,  associated with the non-vanishing entries of the extended Cartan matrix.  Neutral bions are semi-classical realizations of IR-renormalons, and   produce ambiguous imaginary non-perturbative amplitudes semiclassically.    We also show that the  spin wave condensate,   $\langle \partial_{\mu} n^a  \partial_{\mu} n^a  \rangle$, the counter-part of the gluon condensate in gauge theory,   is calculable in the weak coupling regime.  Extending our earlier works on    QCD(adj), deformed Yang-Mills, and  $\C \P^{N-1}$,  we show explicitly that the  ambiguity in the spin wave condensate \cite{David:1983gz}, in the semi-classical regime, receives its dominant  non-perturbative contribution from neutral bions. 
       In the bosonic $O(N\geq 4)$ theory,  we find that fractional kink-instantons generate effects associated  with the strong scale $\Lambda$,  ${\cal K}_j \sim \Lambda^2$   (and not  $\Lambda^{\beta_0} $), and intimately related to the mass gap formation in the theory.  
In 2d theory, despite the triviality of homotopy group, there exists  a non-trivial saddle, solution to the 2d Euclidean equation of motion. We call this 2d-saddle and denote its amplitude  as ${\cal S}_{2d}$.  The action of 2d-saddle is quantized in units of $\frac{4\pi}{g^2}$.  
We show that, in the small-$L$ regime,  the kink-instantons may be viewed as constituents of the 2d-saddle and the two are related via a  Lie algebraic formula:
     \begin{align}
     {\cal S}_{2d} \sim   e^{- \frac{  4 \pi }{g^2 }  } = \Big(\frac{\Lambda}{\mu} \Big)^{2\beta_0} \sim  
 \prod_{j=0}^{\frak r} [{\cal K}_j]^{k_j^\v} , 
  \qquad    
\beta_0= h^\vee= \sum_{i=0}^{{\frak r}  } k_i^\vee, 
\qquad  {\frak r}  = {\rm rank} [\frak  o(N \geq 4)]
  \end{align}
  where ${k_j^\v}$ are the co-marks (dual Kac-labels)  given below in \eqref{co-mark-def}.   The crucial point is that the action of the kink-instantons is $\frac{4\pi}{g^2 \beta_0}$ and survive the large-$N$ limit, unlike the 2d-saddle and 2d-instantons in theories with instantons. 
One may view the vacuum structure of the  $O(N)$ sigma model on $\R\times S^1$ as a    dilute gas of one- and two-events, kink-saddles and bions.  
Our analysis also reveals a surprising degree of similarity between  the classification of twisted classical solutions in the $O(N)$ sigma model on $\R\times S^1_L$  and those in  $O(N)$ gauge theory on $\R^3 \times S^1$ \cite{Argyres:2012ka}.

\subsection{Results for  Grassmannian models}  The vacuum structure of the Grassmannian  sigma models  ${ \rm Gr}(N, M)$   on $\R\times S^1$ can also be described as 
 a    dilute gas of one- and two-events, kink-instantons  and bions, whose actions are $\frac{S_I}{N}$ and $\frac{2S_I}{N}$, respectively,   where $S_I$ is the action of 2d instanton.   
 The 2d-instantons play a relatively minor role  in the bosonic theory, as they are highly suppressed in the semi-classical expansion.  The dynamics of this theory is very similar to the $\C \P^{N-1}$ discussed in our earlier work \cite{Dunne:2012ae}. Extending the analysis of \cite{Dunne:2012ae}, we provided a detailed description of the topological 
 $\Theta$-angle dependence of observables. First, we note that the kink-instanton amplitude is  multi-valued as a function of  the $\Theta$-angle, similar to the monopole operators in deformed-Yang-Mills theory  \cite{Unsal:2012zj,  Anber:2013sga, Bhoonah:2014gpa}  and ${\cal N}=1$ SYM with soft supersymmetry breaking mass term \cite{Poppitz:2012nz,Anber:2014lba}.  We evaluate various observables, such as condensates, and show that the observables are  $2 \pi$ periodic  multi-branched functions as expected by general arguments \cite{Witten:1980sp, Witten:1998uka}.
    In the large-$N$ limit, we explicitly demonstrate the emergence of large-$N$   $\Theta$-angle independence, similar to the Yang-Mills theory \cite{Unsal:2012zj}.

\section{$O(N)$ sigma model in two dimensional spacetime}
\subsection{Basic properties}
The $O(N)$ model is a non-linear sigma model with target space ${\cal T} =S^{N-1}$. We define the theory  on a two-dimensional manifold, $M_2$, and we consider the plane and the cylinder:  $M_2= \R^2$, and $M_2=\R \times \mathbb S^1$.  We consider mostly  the bosonic theory   and  comment briefly on the theory with $N_f$ fermionic species.  $N_f=1$ case is supersymmetric $O(N)$ model. The bosonic field is represented by a real $N$-component unit vector,
$n=(n_1, n_2, \dots n_N)^T$:
\begin{eqnarray}
 n(x)  : \; M_2 \rightarrow  S^{N-1}
 \quad ,  \qquad 
 \sum_{a=1}^{N} n_a^2 (x) =1 
\label{nfield}
\end{eqnarray} 
The classical action of the bosonic model is 
\begin{eqnarray}
S=\frac{1}{2g^2}\int_{M_2}  \left(\partial_\mu n\right)^2 \quad, \quad 
\qquad \frac{1}{g^2} = \frac{N}{\lambda} 
\label{action}
\end{eqnarray} 
where we also defined the 't Hooft coupling $\lambda\equiv N g^2$. The $n$-field is massless  classically and to all orders in perturbation theory for all $N$. Nevertheless, the quantum theory is  believed to be gapped, as can be shown explicitly in the large-$N$ limit  \cite{Novikov:1984rf,Polyakov:1983tt,Gracey:1988us,Hasenfratz:1990zz,abdalla,Vicari:2008jw}.
The quantum theory is  asymptotically free, and has a dynamically generated strong scale, $\Lambda$, given by
  \begin{align}
  \label{RG}
  \Lambda= \mu\, e^{-\frac{2 \pi}{\beta_0\, g^2(\mu)}}\quad , \qquad \beta_0 = N-2
  \end{align}
Here $\mu$ is the UV-cut-off, and $\beta_0$ is the leading coefficient of renormalization group beta-function \cite{Peskin:1995ev, morozov}. Introducing $N_f$ fermions to the $O(N)$ model  (or any other non-linear 2d sigma model) does not alter the leading order beta function, $\beta_0$,  as it  can be deduced by a simple Feynman diagrammatic argument. So, the $O(N)$ model with any number of fermions is always asymptotically free.

The renormalization group $\beta$-function coefficient $\beta_0$  is equal to the dual Coxeter number  $h^\vee $ of the corresponding Lie group, 
 \begin{align}\label{betas}
  \beta_0 =  h^\vee =  \left\{\begin{array} {ll}
  2M-2  = \sum_{i=0}^{M} k_i^\vee &  \quad , \quad  D_M= O(2M)    \cr \cr
  2M-1 =  \sum_{i=0}^{M} k_i^\vee  &  \quad , \quad B_M= O(2M+1) 
  \end{array}
  \right.
  \end{align}
and can be viewed as a sum of  the co-marks or dual Kac labels , $k_i^\vee$ (see, for example, the appendix of  \cite{Argyres:2012ka}). This Lie algebraic interpretation of $\beta_0$ has a natural origin in terms of the compactified saddle solutions, as explained in Section \ref{sec:composite}  below.
The explicit values of the co-marks are: 
\begin{align}
\label{co-mark-def}
D_M= O(2M) : \qquad  (k_0^\vee, \ldots, k_M^\vee)  = (1,1, 2, \ldots, 2, 1,1),   \qquad  M \geq 4 \cr
B_M= O(2M+1) : \qquad    (k_0^\vee, \ldots, k_M^\vee)  =(1,1, 2, \ldots, ,2,1),  \qquad  M \geq 3 
\end{align} 
  The action (\ref{action}) has a  global $SO(N)$ symmetry 
\begin{eqnarray}
n(x)\to \mathcal O\, n(x)\quad,\quad \mathcal O \in SO(N)
\label{global}
\end{eqnarray}
which we later use to impose twisted boundary conditions on $\R \times S^1$. 

The $O(3)$ model has stable instanton solutions on $\R^2$ with action $S_{2d, I} = \frac{4 \pi}{g^2}$.   The base space $\R^2$ combined with a point at $\infty$ can be stereographically projected to a two-sphere, $S^2$, i.e., topologically,  $\R^2 \cup \{\infty\} \sim S^2$.  The $O(3)$ instantons are smooth maps 
$S^2 \rightarrow S^2$, whose degree  takes values in $\pi_2(S^2) = \Z$,  which is the integer-valued topological charge: 
\begin{equation}
Q_T = \frac{1}{8\pi} \int d^2 x \epsilon_{\mu \nu} \epsilon_{abc} 
 \; { n_a}  \partial_\mu  { n_b}   \partial_\nu  {n_c}\quad  \in \Z
\end{equation} 
For $N \geq 4$,  since  $\pi_2(S^{N-1} ) = 0$,  there are no topologically stable instanton configurations according to homotopy theory.   
However, there still exist smooth finite action solutions of the second-order Euclidean classical equations of motion \cite{barbosa,borchers}. Indeed, a simple way to obtain such a solution is to embed an $O(3)$ instanton into $O(N)$ \cite{Din:1979pd}.
 The action is the same as that of the $O(3)$ model instanton, and is quantized. 
  However, such a solution has negative modes when embedded into $O(N)$, indicating that they are saddle points of the action, not minima \cite{Din:1979pd}. 
[Note that these classical Euclidean solutions are not particles -- they cannot decay into something else.]  These solutions were constructed in  \cite{Din:1979pd}, but  their physical significance was not explored: our analysis provides a physical interpretation of these classical saddles.  We will call this object  2d-saddle, and denote its amplitude as  ${\cal S}_{2d}$. In a strict sense, there are no instantons  here, and hopefully, this will eliminate possible  confusions.   The  action and weight associated with these saddles and its relation to 
renormalization group invariant  strong scale $\Lambda$ \eqref{RG} is given by:\footnote{The following observation is true in examples we studied to date. In theories with instantons and a strong scale,  $\Lambda^{ \beta_0} = \mu^{ \beta_0} {\cal I}$. In theories in which there are no instantons,  $\Lambda^{2 \beta_0} = \mu^{2 \beta_0} {\cal S}$,  
where ${\cal S}$ is a saddle with satisfies the second order Euclidean equation of motion.  In $O(3), {\mathbb CP}^{N-1}$,  QCD, SQCD, the first formula is valid  and in $O(N >4)$ and  the Principal Chiral Model (PCM), the second formula holds. ${\cal S}$ saddles are known to have negative modes in their fluctuation operator, and in this  sense, mimic $[{\cal I}   \overline {\cal I} ]$ correlated events. We interpret  ${\cal S}$ as a singularity in the Borel plane, the counterpart of $[{\cal I}   \overline {\cal I} ]$ in bosonic theories with instantons.}
\begin{align}
\label{2d-saddle}
S_{\rm saddle} &= \frac{  4 \pi }{g^2}, \qquad {\cal S}_{2d} \sim e^{- \frac{  4 \pi }{g^2(\mu)}},   
\qquad  \Lambda^{2 \beta_0} = \mu^{2 \beta_0} {\cal S}_{2d}, \qquad   N \geq 4\cr 
S_{\cal I} &= \frac{  4 \pi }{g^2}, \qquad {\cal I}_{2d} \sim e^{- \frac{  4 \pi }{g^2(\mu)}},   
\qquad  \Lambda^{ \beta_0} = \mu^{ \beta_0} {\cal I}_{2d}, \qquad \;\; \;   N =3 
\end{align}

We also consider the base space to be the spatially compactified cylinder, $\R \times \mathbb S^1_L$, and adding the point at infinity,  
$(\R  \cup \{\infty\})  \times \mathbb S^1_L$, we have    $\mathbb S^1 \times \mathbb S^1_L$. This does not ameliorate the situation, since $\pi ( \mathbb S^1 \times \mathbb S^1_L, {S}^{N-1})= 0$ for $N\geq 4$. Therefore,  homotopy considerations suggest that even in the compactified theory there should not be any stable topological defects. However, this is also naive, since in the small-$L$ regime, a potential is induced  on the  target manifold, $ {\cal T} = S^{N-1}$. There are multiple degenerate minima of the potential on  $S^{N-1}$ due to twisted boundary condition, and a potential barrier. In the low energy regime, 
  we will see that there exist a large-class of {\it stable} 1d -instantons, as well as  correlated instanton events.

\subsection{Cartan basis and  twisted boundary conditions} 
 
 In this section we show that the techniques developed for the $\mathbb {CP}^{N-1}$ model \cite{Dunne:2012ae}, can be adapted to the $O(N)$ model, with an additional reality condition on the fields. There are small technical differences between even and odd $N$. We concentrate first on the simply-laced case, $N=2M$, and define the complex parametrization  expressing $S^{2M-1}$ in $\mathbb C^{M}$ rather than in $\R^{2M}$. For each $(2i-1, 2i)$ plane, we define one complex field, instead of two real fields, 
\begin{eqnarray}
&z_i \equiv  n_{2i-1} + i \, n_{2i},  \qquad i=1, \ldots M  
\end{eqnarray}
eigenstates of  rotations  in the $(2i-1, 2i)$-plane $H_{2i-1, 2i}$,  the Cartan-subalgebra generators for $O(2M)$.
The action \eqref{action} can be written as 
\begin{eqnarray}
S= \frac{1}{2g^2}\int_{M_2}  \left| \partial_\mu  z_i \right|^2, \qquad  \sum_{i=1}^{M} | z_i(x)|^2 =1
\label{action2}
\end{eqnarray} 
This form only makes  the $SU(M) \times U(1)$ subgroup of $SO(2M)$ manifest, but the full symmetry is  still present.

{\bf  The rationale behind twisted boundary conditions:}
As in the $\mathbb {CP}^{N-1}$ model \cite{Dunne:2012ae} and $U(N)$ Principal Chiral Model \cite{Cherman:2013yfa}, we define suitable twisted boundary conditions for which the twisted free energy density  scales with $N$ as $O(N^0)\,T^2$, as opposed to $O(N^1)\,T^2$.    
 The rationale behind this is  to have the weak-coupling semiclassical theory on the compactified cylinder continuously and adiabatically connected to the gapped $\R^2$ theory, which would not be possible if  the free energy scaled as $O(N^1)\,T^2$, as in the deconfined regime where the gapless microscopic bosonic and fermionic degrees of freedom are liberated.
Our goal is to capture a regime which resembles  the theory on $\R^2$ as closely as possible. The details of this rationale can be found in \cite{Dunne:2012ae,Cherman:2013yfa}. 
  
In terms of the Cartan  matrix  eigenstates $z_i$ (and their fermionic partners), the twisted boundary conditions amounts to 
 \begin{align} 
&  \begin{pmatrix}
 (z_i)  \cr
 (z_i)^{*}  
\end{pmatrix}   (x_1, x_2+L)     = 
  \begin{pmatrix}
\Omega_0 &  \cr
 & \Omega_0^{\dagger} \cr
\end{pmatrix}  
 \begin{pmatrix}
 (z_i)  \cr
 (z_i)^{*}  
\end{pmatrix}   (x_1, x_2)    
\qquad {\rm for} \;\;  O(2M)    \cr \cr
& \begin{pmatrix}
 (z_i)  \cr
 (z_i)^{*}   \cr
 n_{2M+1} 
\end{pmatrix}   (x_1, x_2+L)     = 
  \begin{pmatrix}
\Omega_0 &  & \cr
& \Omega_0^{\dagger} &  \cr 
&&1
\end{pmatrix}  
 \begin{pmatrix}
 (z_i)  \cr
 (z_i)^{*}  \cr
  n_{2M+1} 
\end{pmatrix}   (x_1, x_2)    
\label{twist2} \qquad {\rm for} \;\;  O(2M+1)   
\end{align}
where $L$ is the spatial compactification scale, and the twist matrix is
\begin{eqnarray}
\label{twistmatrix}
\Omega_0&=&\begin{pmatrix}
e^{ 2 \pi i  \mu_1}  &0&\dots &0 \cr
0& e^{2 \pi i  \mu_2}   &\dots &0\cr
\vdots\cr
0&0&\dots & e^{2 \pi i   \mu_M} 
\end{pmatrix} 
\quad, \quad 0\leq \mu_M \leq \mu_{M-1} \leq \dots \leq\mu_1 \leq   \half
\end{eqnarray}
The restriction $0\leq \mu_j\leq \frac{1}{2}$ arises from the reality of the original $O(N)$  fields. Note that for $O(2M)$, all eigenvalues are paired, while for $O(2M+1)$, only one of the eigenvalues of the  logarithm  of twist matrix  is actually zero, and unpaired. 

\begin{figure}[ht]
\begin{center}
\includegraphics[angle=0, width=6.in]{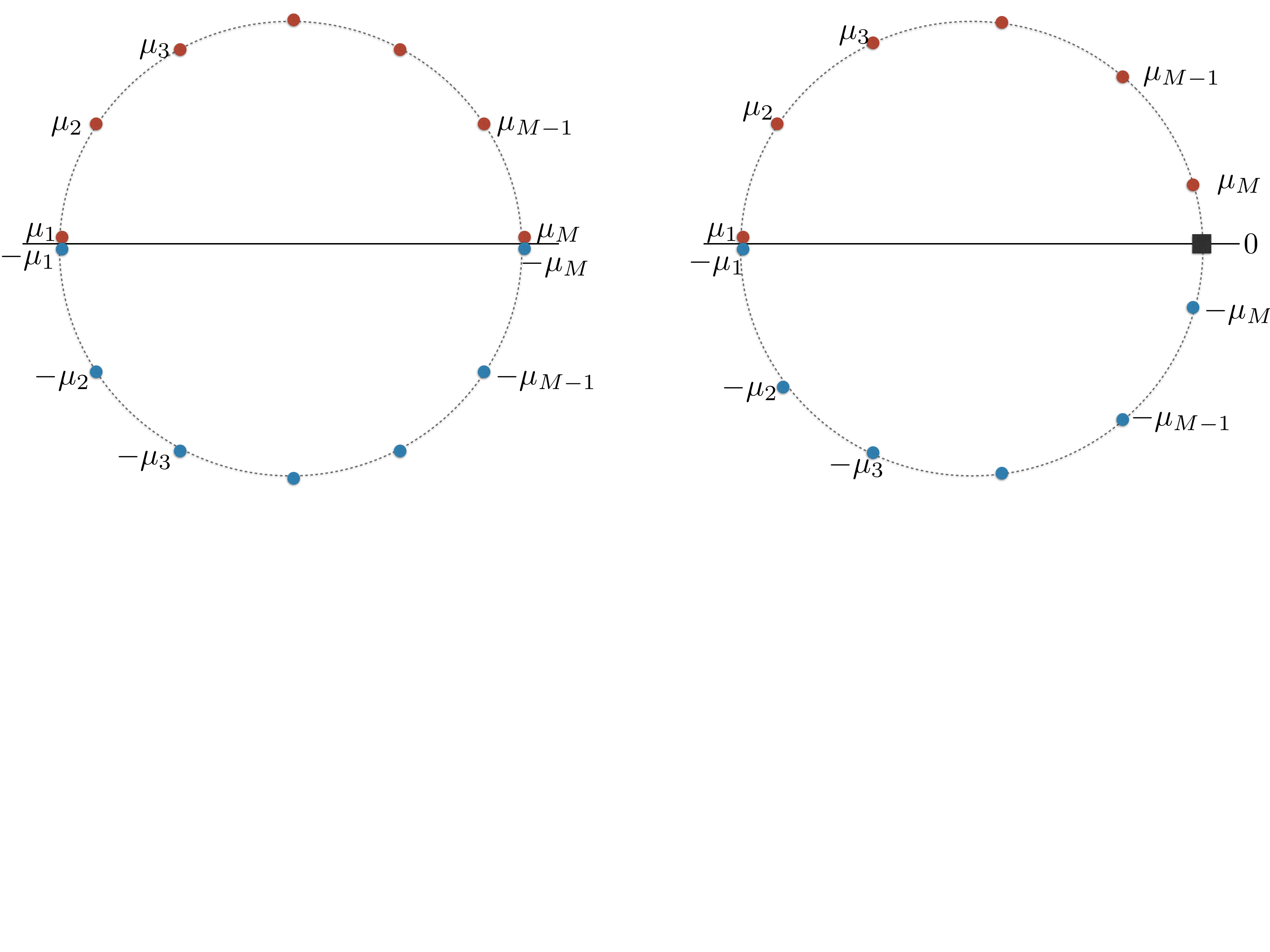}
\vspace{-5.5cm}
\caption{Twisted boundary condition for the $O(2M)$ (left)   and  $O(2M+1)$ (right)  models. The blue points are mirror images of the red ones.   
$\pm \mu_1$ and $\pm\mu_M$  are coincident, but are split  for convenience of visualization. 
For $O (2M+1)$,  the eigenvalue at zero  does not have a mirror image.  }
\label {fig:TBC}
\end{center}
\end{figure}

These boundary conditions  can be traded for a background field for the complex  $\arg(z_i)$ field, similar to $\mathbb {CP}^{M-1}$ \cite{Dunne:2012ae}.
  A field redefinition introduces periodic fields (with tilde on them)  and a  background  ``holonomy". 
\begin{eqnarray}
\tilde z_i (x_1, x_2)= e^{ - i 2\pi  \mu_i x_2/L}  z_i  (x_1, x_2), \qquad  
\tilde z_i (x_1, x_2+L)= \tilde z_i  (x_1, x_2)\qquad , \qquad  
\label{twist3}
\end{eqnarray}
where  the action takes the form 
\begin{eqnarray}
S= \frac{1}{2g^2}\int_{\R \times S^1_L}  \left| D_\mu \tilde z_i \right|^2  , \qquad D_\mu \tilde z_i = \left( \partial_\mu  + i \delta_{\mu 2} \frac{2\pi \mu_i}{L} \right)  \tilde z_i
\label{action-2}
\end{eqnarray} 
We choose the boundary conditions for which the twist free energy is minimal  \cite{Dunne:2012ae,Cherman:2013yfa}.  This amounts to the choice depicted in Fig.\ref{fig:TBC}, for which the $M$ eigenvalues are distributed as uniformly as possible in the interval $[0, \pi]$:
\begin{eqnarray}
\Big( \mu_M,  \mu_{M-1},  \ldots, \mu_2,  \mu_1\Big) =  
\Big( 0,  \tfrac{1}{2(M-1)},  \tfrac{2}{2(M-1)}, \ldots,   \tfrac{M-2}{2(M-1)}, \tfrac{1}{2}  \Big)
 \label{background-2}
\end{eqnarray}
Notice that the eigenvalues appear in $\pm$ pairs, due to the reality condition on the fields. Thus the interval $[0, \pi]$ is divided into $M-1$ equal wedges, rather than $M$, thus
the  wedges have angle $\frac{2\pi}{2M-2}$.

For odd $N$, the situation is similar, with a small technical difference that mimics the different root structure of the associated Lie algebra, since $O(2M+1)$ is non-simply laced. Take $N=2M+1$.  
The twist matrix $\Omega_0$ is again an $M\times M$ matrix, but with 
\begin{eqnarray}
\Big( \mu_M,   \mu_{M-1}, \ldots, \mu_2, \mu_1  \Big) =   
 \Big(  \tfrac{1}{2}  \tfrac{1 }{2M-1}, \tfrac{1}{2}  \tfrac{3 }{2M-1},   \tfrac{1}{2}  \tfrac{5 }{2M-1},   \ldots,   \tfrac{1}{2}  \tfrac{2M-3 }{2M-1},   \tfrac{1}{2} \Big)
\end{eqnarray}
This accounts for the non-simply-laced structure of the algebra, and is represented in Fig. \ref{fig:TBC}(right). Note that the eigenvalue  at zero 
is not paired, and is not dynamical.  The long  wedges have angle $\frac{2\pi}{2M-1}$, and the short wedge  (between zero and $\mu_M$ ) has angle  $\frac{\pi}{2M-1}$. 

%
\subsection{Embedding $O(3)$ into $O(N)$}
 
 To construct fundamental classical solutions in $O(2M)$ with twisted boundary conditions, we embed
 $O(3)$ solutions into   $O(2M)$. This is analogous to the construction of  topological instanton or monopole-instanton  solutions in gauge theory, built out of $SU(2)$ solutions embedded into the larger  gauge group, e.g. into $SU(N)$. 
 The main idea is that  $O(3)$, similar to $SU(2)$, is rank-1, and is the minimal possible group structure in which a non-trivial solution can live. 
 
To this end, we   work within the subspaces respecting the appropriate constraints for the $z(x_1, x_2)$ field.  
So, 
 consider only two out of $M$-components turned on: 
 \begin{equation} 
  \begin{pmatrix}
z_1 \cr
z_2 \cr
\vdots\cr
z_{j+1} \cr
z_{j+2}  \cr
\vdots\cr
z_M \cr
\end{pmatrix}     \longrightarrow 
  \begin{pmatrix}
0 \cr
0 \cr
\vdots\cr
z_{j+1} \cr
z_{j+2}  \cr
\vdots\cr
0 \cr
\end{pmatrix}  \qquad  {\rm  such \; that } \;\;  |z_{j+1}|^2 + |z_{j+2}|^2=1
\end{equation}
However, the 
 $ (z_{j+1}, z_{j+2})^T  $
is an $O(4)$ or rank-2 object, with 3 real degrees of freedom, whereas an $O(3)$  solution should be a  rank-1 structure, with 2 degrees of freedom.

We can have an $O(3) \sim \mathbb {CP}^1 $ living in $O(4)$ by using a parametrization that is most convenient for   using the twisted boundary conditions: 
 \begin{equation}
\label{doublet}
  \begin{pmatrix}
\frak z_{j+1}\cr
\frak z_{j+2}  
\end{pmatrix} =
   \begin{pmatrix}
e^{i  \phi/2}  \cos \frac{\theta}{2} \cr
e^{-i  \phi/2 }   \sin \frac{\theta}{2} 
\end{pmatrix}   
\end{equation}
 Here, the coordinates  
$( \frak z_{j+1}, 
\frak  z_{j+2})^{T} $ can be obtained by gauging an overall $U(1)$ factor  in  the target space $S^3$ of the $O(4)$ model.  
 The
$\mathbb C \mathbb P^{\rm 1}$ model is equivalent to the $O(3)$  non-linear $\sigma$-model through 
the  simple identification of fields:   
\begin{equation}
\vec n(x) = \frak z_{j+a}^\dagger \vec \sigma_{ab} \frak z_{j+b}\qquad   
\label{map-2}
\end{equation}
where   $\vec \sigma$ are the Pauli matrices.

Now we see the effect of the twisted boundary conditions (\ref{twist2}): 
\begin{equation}
\label{twisted-doublet}
  \begin{pmatrix}
\frak z_{j+1} \cr
\frak z_{j+2}  
\end{pmatrix}  (x_1, x_2+L) =   \begin{pmatrix}
e^{ i 2 \pi \mu_{j+1}}  \frak z_{j+1} \cr
e^{ i 2 \pi \mu_{j+2}} \frak z_{j+2}  
\end{pmatrix}  (x_1, x_2) 
\end{equation} 
One can undo the twist in favor of a background field and periodic fields  $(\theta, \phi) \in S^2$. Using 
\eqref{map-2}, this amounts to the following modification of the $S^2$ coordinates, in terms of original real fields $n_a$ that we have used:  
\begin{eqnarray}
 \left ( \begin{array}{l}
n_1 \cr
 n_2 \cr
 n_3
  \end{array} \right)  = \left ( \begin{array}{l}
 \sin \theta \cos \left(\phi +  \xi x_2 \right)   \cr
 \sin \theta \sin \left(\phi +  \xi x_2 \right)
    \cr \cos \theta  
  \end{array} \right)  \, , 
\quad   \qquad \xi \equiv  \frac{2\pi}{L}   \alpha_{j+1} . \mu 
      \label{par2}
\end{eqnarray}
The twisted background emulates  a fractional momentum insertion in the compact  $x_2$ direction, and this has interesting consequences.  
The resulting  Lagrangian  on ${\R \times \mathbb S^1}$  is given by  
\begin{equation}
 S= 
\frac{1}{2 g^2} \int_{\R \times  \mathbb S^1_L}  (\partial_\mu \theta)^2 + 
\sin^2 \theta   (\partial_\mu \phi + \xi \delta_{\mu2})^2  
\label{action3}
\end{equation}

\subsection{One-events: Kink-saddles} 
In the small-$L$  semiclassical limit, the 2d QFT reduces to an effective quantum mechanics (QM) problem, by a Kaluza-Klein decomposition of the fields  \cite{Dunne:2012ae,Cherman:2013yfa}.
As $L\to 0$, the reduced QM action becomes
 \begin{eqnarray}
 S    & =& \frac{1}{2g^2}\int_{\R }  \Big[ ( \partial_t  \theta )^2 + \sin^2 \theta  ( \partial_t  \phi_1 )^2 +   \xi^2 \sin^2 \theta       \Big] 
  \label{actionQM} \end{eqnarray}
  where the effect of the twist-term is to create a potential barrier between the north and south pole of the two-sphere, $S^2$.
 The Hamiltonian  is given by, 
\begin{eqnarray}
H&&= \frac{g^2}{2} P_\theta^2  + \frac{g^2 }{2  \sin^2 \theta} P_{\phi_1}^2  +
 \frac{ \xi^2}{2g^2} \sin^2 \theta    
   \label{Ham-1}
\end{eqnarray} 
The $\phi$-fluctuations, when quantized, are gapped. Since $\phi$ is cyclic coordinate, the Hamiltinian is already diagonal in the angular momentum $P_{\phi}$-basis, with quantum numbers 
$m_{\phi}=0, \pm 1, \pm 2, \ldots $. Since the gap in the $\phi$-sector is of the order  $g^2/L$,  and the low energy physics is governed by  non-perturbatively small energy splitting,  we set  $m_{\phi}=0$ from here on. This is the justification of the Born-Oppenheimer approximation. Thus, it suffices to study  the action
 \begin{eqnarray}
 S    & =& \frac{1}{2g^2}\int_{\R }     \Big[  ( \partial_t  \theta )^2 + \xi^2 \sin^2 \theta       \Big], 
  \label{actionQM-2} \end{eqnarray}
except for the study of affine-kink saddle (and related ones) to be discussed below. Explicit solutions are expressed in terms of Sine-Gordon kinks, as in the $\C\P^{N-1}$ model  \cite{Dunne:2012ae}.  

\begin{figure}[ht]
\begin{center}
\includegraphics[angle=0, width=5.in]{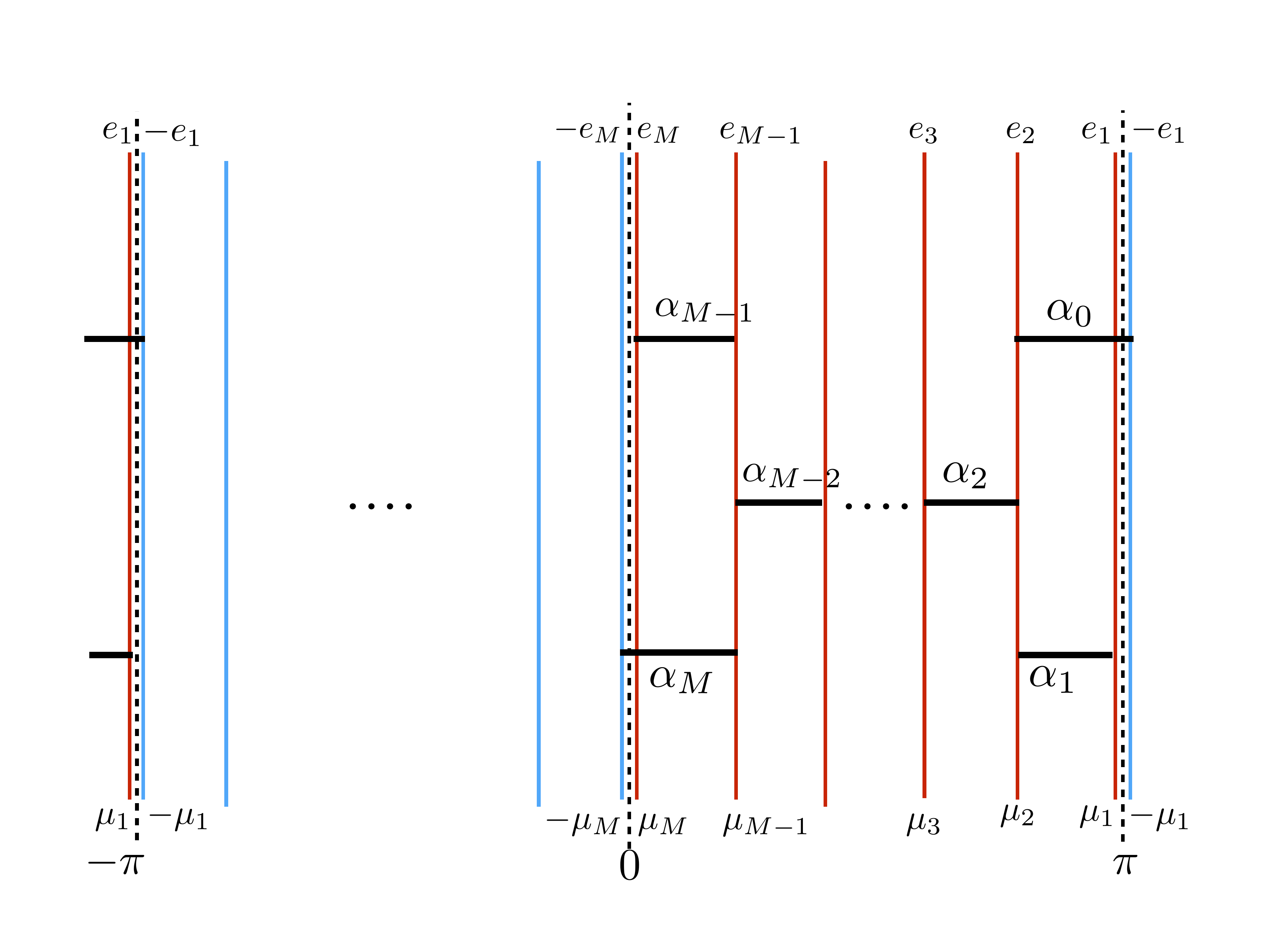}
\includegraphics[angle=0, width=5.in]{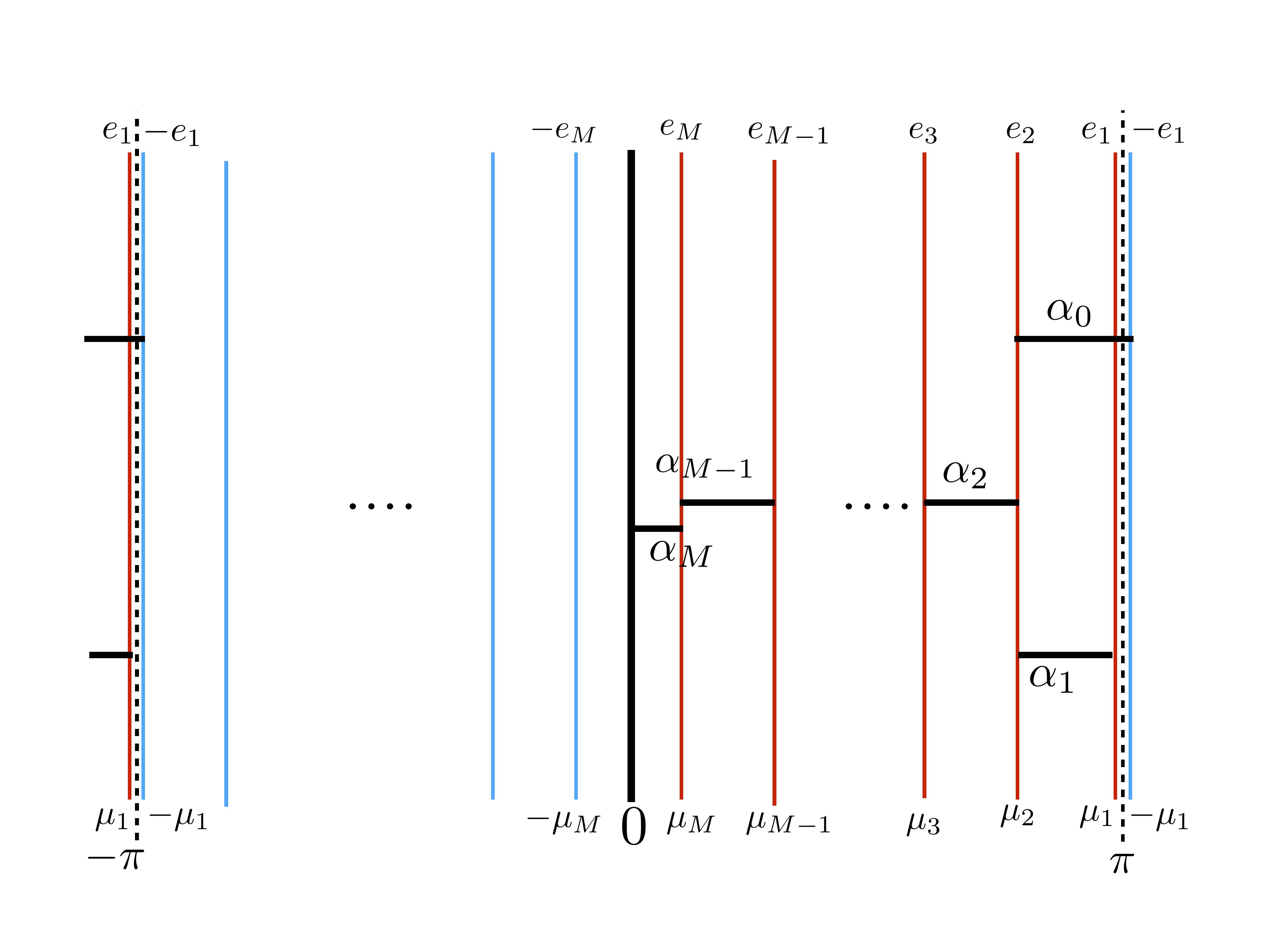}
\vspace{-1.3cm}
\caption{ Minimal action saddles  associated with the affine root system of the simply-laced  ${\frak o} (2M)$ algebra (top)  and the root system of  the non-simply laced ${\frak o} (2M+1)$ algebra (bottom). 
In the simply laced  case,  the action of the saddles are proportional to the eigenvalue differences. In the non-simply laced case, the short root (and its KK-tower) requires more care, as discussed in the text.}
\label {fig:TBC-instanton}
\end{center}
\end{figure}
\begin{figure}[ht]
\begin{center}
\includegraphics[angle=0, width=5in]{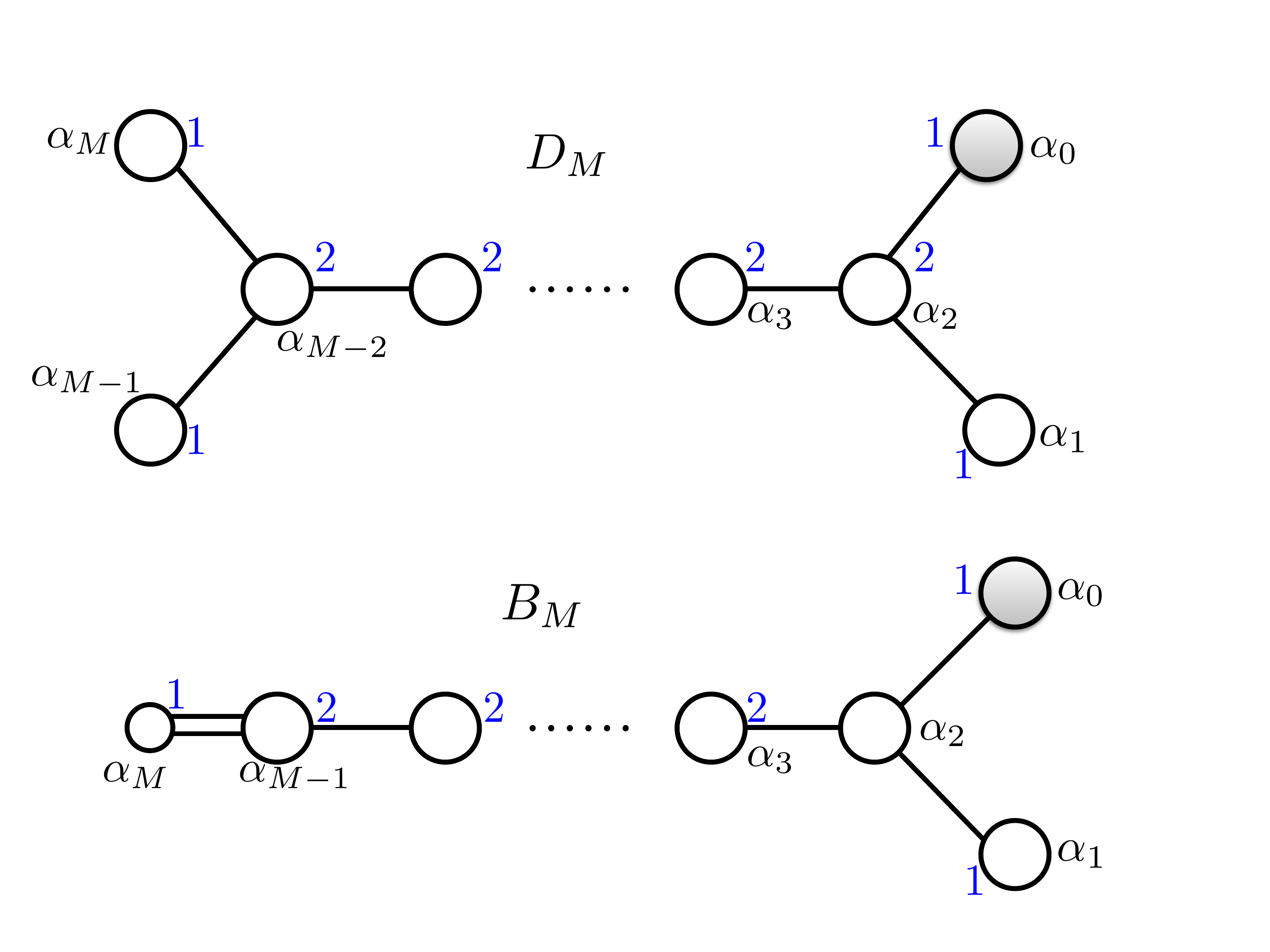}
\vspace{-0.5cm}
\caption{ Fig.~\ref{fig:TBC-instanton} can be reinterpreted in terms of the Dynkin diagrams of   $D_M=O(2M)$ and  $B_M=O(2M+1)$. The shaded circles  denote the  affine affine roots, and are present because the theory is compactified on a circle. There is a short root in the non-simply-laced $O(2M+1)$ case.   The above diagrams  should be used for $B_{M \geq 3}=O((2M+1) \geq 7) $, and  $D_{M \geq 4}=O((2M) \geq 8)$, with lower rank cases requiring slightly more care due to additional symmetries. 
}
\label {fig:dynkin}
\end{center}
\end{figure}

{\bf Simply laced  ${\frak o}(2M)$ case:}
The kink-saddles are associated with the roots of the Lie algebra. Again we first illustrate with the simply-laced $O(2M)$. The minimal action saddles are in one-to-one correspondence with the affine root system
\begin{align}
\Delta_{\rm aff} =\Big \{ \alpha_i =e_i - e_{i+1}, \;  \alpha_M =e_{M-1} + e_{M},  \;  \alpha_{0} =-e_{1} - e_{2} \Big\}
\end{align}
where $ i=1, \ldots, M-1$.  
These  saddles are associated with the tunneling in field space between the following configurations: 
\begin{align} 
&
{\cal K}_1:   \begin{pmatrix}
1 \cr
0 \cr
\vdots\cr
0 \cr
0 \cr
\end{pmatrix}    \longrightarrow
  \begin{pmatrix}
0 \cr
1 \cr
\vdots\cr
0 \cr
0 \cr
\end{pmatrix} , \qquad 
{\cal K}_2:   \begin{pmatrix}
0 \cr
1 \cr
0 \cr
\vdots\cr
0 \cr
\end{pmatrix}    \longrightarrow
  \begin{pmatrix}
0 \cr
0 \cr
1 \cr
\vdots\cr
0 \cr
\end{pmatrix},   \ldots, 
{\cal K}_{M-1}:   \begin{pmatrix}
0 \cr
0 \cr
\vdots\cr
1 \cr
0 \cr
\end{pmatrix}    \longrightarrow
  \begin{pmatrix}
0 \cr
0 \cr
\vdots\cr
0 \cr
1 \cr
\end{pmatrix}  \cr
&{\cal K}_{M}:   \begin{pmatrix}
0 \cr
0 \cr
\vdots\cr
1 \cr
0 \cr
\end{pmatrix}    \longrightarrow
  \begin{pmatrix}
0 \cr
0 \cr
\vdots\cr
0 \cr
-1 \cr
\end{pmatrix}  \qquad 
{\cal K}_0:   \begin{pmatrix}
-1 \cr
0 \cr
\vdots\cr
0 \cr
0 \cr
\end{pmatrix}    \longrightarrow
  \begin{pmatrix}
0 \cr
1 \cr
\vdots\cr
0 \cr
0 \cr
\end{pmatrix} , \qquad 
\label{instantons2M}
\end{align}
Since $O(2M)$ is simply laced, the actions of the 1d-saddles are proportional to the distance between the eigenvalues of  $-i \log \Omega$. 
With our choice of ordering of $\mu_i$, 
the action of the 1d-saddle events for $i=0, 1, \ldots, M$ is given by 
\begin{align}
\label{actions-0}
S_{{\cal K}_{_i}} \equiv   S_0 = \frac{2 \xi}{g^2} = \frac{  4 \pi}{g^2}   ( \bar \alpha_i. \mu)  =  \frac{  4 \pi }{g^2 (2M-2)}=  \frac{  4 \pi }{g^2\beta_0} 
\end{align}
where we used special notation (following \cite{Argyres:2012ka}) 
\begin{align}\label{}
\ba_j. \mu  := \a_j.\mu+ \d_{j,0}  = \left\{ \begin{array}{ll}
(\a_j .\mu),   &  j=1, \ldots, M \cr
 (\a_0.\mu) +1,  &  j=0\cr
 \end{array} \right\}
\end{align}
This means, special care is needed for the affine root, $\a_0$.  One may be tempted to think that the action for the  $\a_0$ kink-saddle is  the absolute value of $\frac{  4 \pi  (\mu_1 + \mu_2)}{g^2}=\frac{  4 \pi  (2M-3)}{g^2(2M-2)} $, which is $(2M-3)$ times larger than the others.  However, this is not the least action configuration associated with 
$\Delta z \; = \alpha_0 $ tunneling event.   In fact, recall that 
the twisted background behaves as a fractional momentum insertion in the compact direction. Combined with  $n=1$ units of 
Kaluza-Klein momentum for the $\phi$ field,  and then, performing a ``twisted" dimensional reduction,   one obtains 
 \begin{eqnarray}
S&&= \frac{1}{2g^2}\int_{\R }  \Big[ ( \partial_t  \theta )^2 +   (2 \pi (1+ \a_0.\mu))^2 \sin^2 \theta       \Big] , 
  \label{actionQM3}
\end{eqnarray} 
and the   minimal action of the tunneling event associated with $\alpha_0$ is given by \eqref{actions-0}. 
This classical solution is  the counter-part of the twisted monopole-instanton in gauge theory on $\R^3 \times S^1$ \cite{Lee:1997vp,Kraan:1998pm,Argyres:2012ka}.

{\bf Non-simply laced  ${\frak o}(2M+1)$ case:}  
The minimal action saddles are in one-to-one correspondence with the affine root system
\begin{align}
\Delta_{\rm aff} =\Big \{ \alpha_i =e_i - e_{i+1}, \;  \alpha_M = e_{M},  \;  \alpha_{0} =-e_{1} - e_{2}  \Big\}
\end{align}
where $ i=1, \ldots, M-1$.  
Note that all roots but $\alpha_M$ has length square-root two.  The length of $\alpha_M$ is  just one.   This has a small effect in determination of the action of 
kink-instantons.    \eqref{actions-0} used in the action of the kink-instantons  is  valid only for simply-laced case. 
A formula for the kink-instanton action valid for both simply and non-simply laced algebras is given by:
\begin{align}
S_{{\cal K}_{_i}}  = \frac{  4 \pi}{g^2}   \left(  \frac{2  (\ba_j .\mu) }{\a_j.\a_j}   \right) =  \frac{  4 \pi}{g^2} \left\{ \begin{array}{ll}
(\a_j .\mu),   &  j=1, \ldots, M-1 \cr
 (\a_0.\mu) +1,  &  j=0\cr
 2  (\a_M.\mu),  &     j=M
\end{array} \right\}
=  \frac{  4\pi }{g^2(2M-1)}=  \frac{  4 \pi }{g^2\beta_0} 
\end{align}
i.e, there is an extra factor of two for the short root. 
In the penultimate step, we used the background \eqref{background-2} where the actions of all the kink-instantons become equal.

\subsection{Two-events: Charged and neutral bions}
The classification of bions, the correlated two-events,   is identical to the $\mathbb {CP}^{N-1} $ model.  Two defects are universal and are in one-to-one correspondence with the non-vanishing entries of the extended Cartan matrix. So, the only difference with respect to the  $\mathbb {CP}^{N-1} $ case is the replacement of the $SU(N)$ extended Cartan matrix with the $O(N)$ one. For the discussion of the correlated amplitudes and derivations, see \cite{Dunne:2012ae}. 
  \begin{itemize}
\item  {\it Charged bions:} For each  non-vanishing negative entry of the extended Cartan matrix,  $\hat A_{ij}<0$,  there exists a bion ${\cal B}_{ij} = [ {\K}_i \bar\K_j] \sim e^{-2S_0} $,  associated with the tunneling event 
\be 
\tilde z  \longrightarrow \tilde  z + \alpha_i - \alpha_j    \qquad  \alpha_i \in \G_r^\v
\ee
\item {\it  Neutral bions:}  For each  non-vanishing positive entry of the extended Cartan matrix,    $\hat A_{ii}>0$,  and     there exists a bion ${\cal B}_{ii, \pm} = [\K_i\bar\K_i]_{\pm}$ with  vanishing topological charge and  associated with the tunneling-anti-tunneling  event 
\be 
\tilde z  \longrightarrow \tilde  z   + \alpha_i - \alpha_i   \qquad  \alpha_i \in \G_r^\v
\ee
The  neutral bion amplitude is two-fold ambiguous in the bosonic model. 
  \end{itemize}

\subsection{2-d Saddles as a composite at long distances}
\label{sec:composite}

In  2-dimensions, instantons are exact BPS solutions for the $O(3)$ model, but for $O(N)$ with $N\geq 4$, they satisfy the second-order Euclidean equations of motion, 
and possess  negative modes in the fluctuation operator around them. Despite this,  the action of these configurations is  quantized in units of instanton action. 
These are harmonic maps, and as such, they are finite action extrema of the given action functional.  To distinguish from instantons which are solutions to first order equations, 
we will refer to these as saddles. They are the analog of   the ``unitons'' of the  Principal Chiral Model \cite{uhlenbeck,Ward:1990vc,Dunne:1992hq,Cherman:2013yfa}:\footnote{In the case of $O(4)$ model, target space is $S^3$, same as the $SU(2)$ PCM. In that case, the $O(4)$ saddle is exactly same as $SU(2)$ uniton.} 
\begin{align}
S_{\rm saddle} = \frac{  4 \pi }{g^2} 
\end{align}
The 1d-kink saddles of the previous section may be viewed as the constituents of these 2d saddles. The relation is Lie algebraic, and applies to both $O(2M)$ and $O(2M+1)$. 
 There exists a unique positive integral linear relation among the simple and affine co-roots, 
 \begin{equation} 
 \sum_{j=0}^{ \frak r} k_j^\v \a_j^\v = 0,  \qquad  {\frak r}  = {\rm rank} [\frak  o(N)]
 \end{equation}
 with $k_0^\v=1$, where     $k_j^\v$ are called the dual Kac labels (or co-marks).   
 This mathematical relation defines  the physical relation between the 2d saddle  and its constituent kink-saddles. 
Combining $k_j^\v$ kink-saddles of type ${\cal K}_j$ for $j=0,\ldots, {\frak r}$ we obtain the amplitude:
\begin{align}
 {\cal S}_{2d}  \sim  \prod_{j=0}^{ \frak r} [{\cal K}_j]^{k_j^\v} = e^{- \frac{  4 \pi }{g^2\beta_0} \sum_{j=0}^{\frak r} k_j^\v   } =  e^{- \frac{  4 \pi }{g^2\beta_0}  h^\v } =  e^{- \frac{  4 \pi }{g^2}  }  
  \end{align}
In the last step, we used the fact that   the beta-function coefficient $\beta_0$ is exactly equal to the   dual Coxeter number   $h^\v$. 
 Thus,  the 2d saddle may be viewed, at least in the weak coupling regime, as fractionalizing into   ${\frak r}+1$   kink-saddles with multiplicities equal to the co-marks, $k_j^\v$.

\subsection{Euclidean  description of the vacuum}
\begin{figure}[ht]
\begin{center}
\includegraphics[angle=0, width=6in]{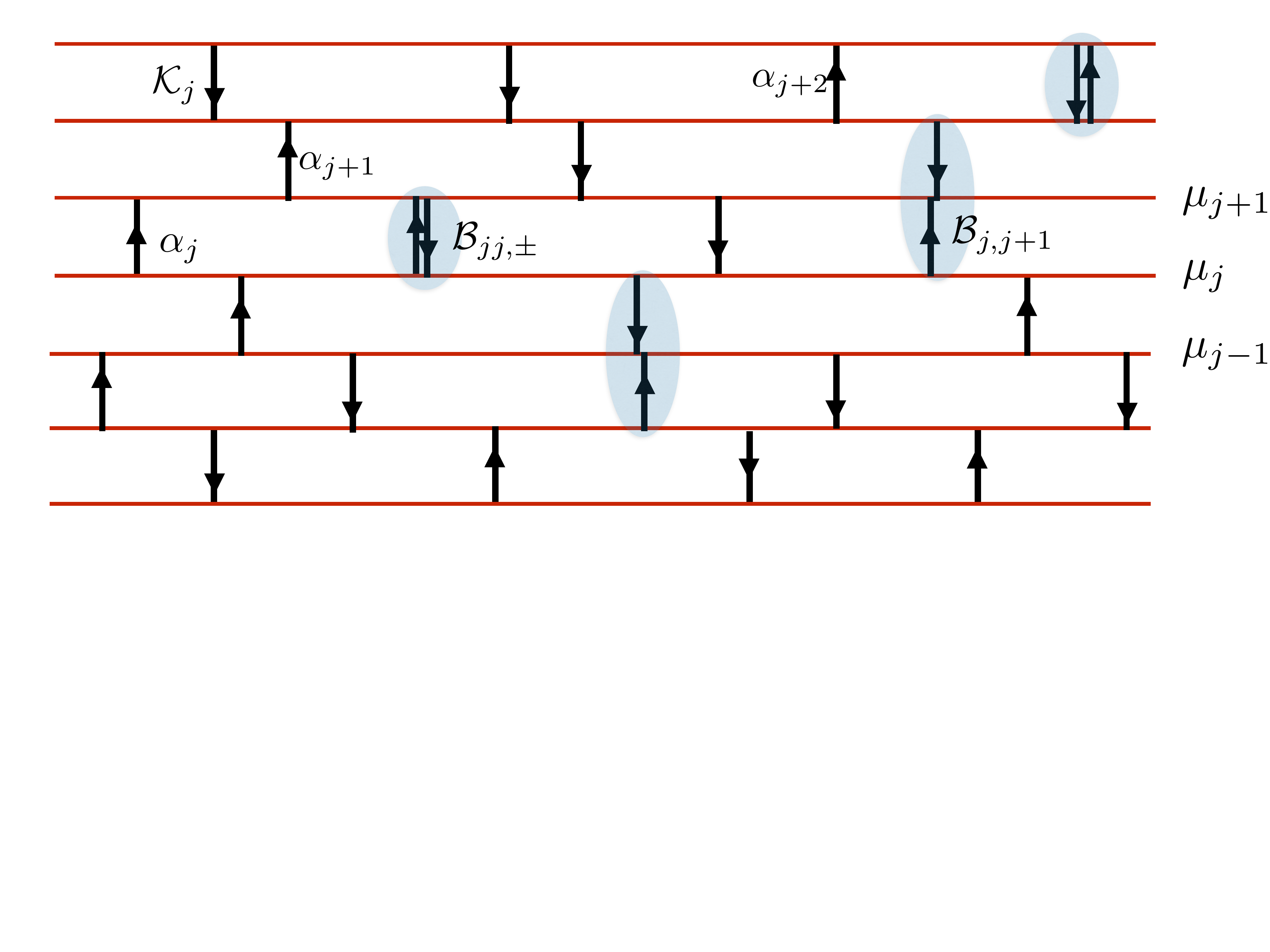}
\vspace{-5.5cm}
\caption{A snap-shot of the  dilute gas of one- and two-events, kink-saddles (associated with roots) and bions (associated with non-zero entries of the extended Cartan matrix), respectively.  The amplitude of the neutral bions are two-fold ambiguous, fixing the ambiguity of perturbation theory. 
 }
\label {fig:dkbg}
\end{center}
\end{figure}

  The Euclidean vacuum of the $O(N)$ model may be viewed as a dilute gas of one-, two-, {\it etc} events, a snap-shot of which is depicted in  Fig.\ref{fig:dkbg}.  
  The density of the k-event is $e^{-k \frac{4\pi}{g^2 \beta_0} }$, thus, the densities are hierarchical. Most of the interesting non-perturbative phenomena are sourced by one and two-events, e.g, mass gap,  semi-classical realization of renormalons.

The small-$LN\Lambda$ theory, constructed by using twisted boundary conditions  {\it remembers}  the strong scale  and dynamics of the two-dimensional theory to a large-degree.  The non-perturbative gap in the spectrum of the theory in the regime $L N \Lambda \lesssim 1$ is a kink-saddle effect and is given by 
\begin{align}
m_g \sim (LN)^{-1} e^{-S_{{\cal K}_{\alpha_i}}} =(LN)^{-1}  e^{- \frac{  4 \pi }{g^2\beta_0} } =   \Lambda (\Lambda L N)
\end{align}
The weak coupling semi-classical approximation breaks down at $LN \Lambda \sim 1$, where 
the semi-classical gap reaches  $  \sim \Lambda$.   One may surmise that the gap is saturated at this scale for  $L N \Lambda \gtrsim 1$. 
At the two-event level, the neutral bion is two-fold ambiguous, and this ambiguity cancels exactly the ambiguity of perturbation theory in the small-$S^1$ regime. 

We would like to comment briefly why this picture is relatively surprising. Historically, $O(N>3)$ model is viewed as a theory with no instantons. Sometimes, it is also erroneously asserted that  $O(N)$ model is a field theory with  a mass gap, but with no non-perturbative saddles. 
Both  perspectives are often presented as a point of divergence from the interesting 4d gauge theories. Consequently, more interest is given to theories with instantons.  The picture that emerges by combining  resurgence and continuity instructs us that in $O(N)$ model, the classification of saddles in $O(N)$ model on $\R \times S^1$  is essentially isomorphic to the one of gauge theory on $\R^3 \times S^1$.

\subsection{Semi-classical realization of  IR-renormalons as neutral bions} 
 The operator product expansion (OPE) connects perturbative information with non-perturbative condensates. The rate of the  factorial divergences associated with infrared (IR) renormalons 
 are identified with   certain condensates of specific dimensions  \cite{thooft,parisi,mueller,beneke,shifman-review}. 
 On $\R^2$,   the exact   large-$N$ solution provides a rigorous realization of this idea.  The first IR renormalon Borel singularity  in $O(N)$ model   and the OPE spin-wave condensate are related as   \cite{David} 
\begin{eqnarray}
t_{\rm IR}^{\R^2}=2 \times \frac{4 \pi}{\beta_0\, g^2(Q^2)} \leftrightarrow    \langle   { O}_1  \rangle  =\langle \partial_{\mu} n^a  \partial_{\mu} n^a  \rangle  \leftrightarrow \left(\frac{\Lambda^2}{Q^2}\right) \,,
\label{one}
\end{eqnarray}
where $\beta_0$ is the first coefficient of the $\beta$ function. Equivalently, one can state that the ${ O}_1$  vev is  two-fold ambiguous: Let 
 $\theta=\arg(g^2)$.
 \begin{align}
 \langle{ O}_1 \rangle_{\theta= 0^{\pm}} = c_1\Lambda^2 \pm i d_1 \Lambda^2 \; , 
  \end{align}
 This ambiguity arises because $\arg(g^2)=0$ is a Stokes lines. This ambiguity cancels the ambiguity in the Borel resummation of perturbation theory along the Stokes line. 
On $\R^2$ there is no known semi-classical  understanding of this ambiguity.  
 
 Note that the 2d-saddle we discussed around \eqref{2d-saddle} may have negative modes and may also have ambiguities. This is morally similar to the instanton-anti-instanton ambiguity in the theories with instantons. This singularity, similar to the instanton-anti-instanton singularity,  is rather far from the origin of the Borel plane, it can only give a very suppressed power law correction especially at large-$N$, 
$ {\cal S}_{2d} \sim \left(\frac{\Lambda^2}{Q^2}\right)^{\beta_0} $. Clearly, the renormalon singularity   discussed above is approximately $N$-times closer to the origin.

Our semi-classical analysis of the $O(N)$ model, using spatial compactification to the cylinder $\R\times S^1_L$ along with adiabatic continuity, provides a semi-classical realization  of the leading IR-renormalon in the small-circle regime.   Due to asymptotic freedom, in this weak-coupling semi-classical domain we can
calculate the  non-perturbative contribution to  $ \langle   { O}_1  \rangle $.  In the Euclidean path integral representation, the  condensate  $ \langle   { O}_1  \rangle$ receives its leading non-perturbative contribution from the kink-saddles.   Fig.~\ref{fig:dkbg} represents a snap-shot of the Euclidean vacuum of the theory. It is, as described already, a dilute gas of one-events, two-events, three-events etc.  

Calculating the  vacuum expectation value $O_1$   at leading order in semi-classics   is equivalent to finding the average of  the  action density  $\half \partial_{\mu} n^a  \partial_{\mu} n^a $ over all space.  The action density  is concentrated  within the characteristic size $r_\K $  of the kink-saddles.   The condensate, at leading semi-classical order, is therefore proportional to the density of the kink-saddle events. Both kink-saddles  and  anti-kink saddles contribute to 
$O_1 $ in the same manner 
 $ \langle   { O}_1  \rangle   \propto    \sum_{j}  S_{{\cal K}_{j}}   \left( {\cal K}_{j}   + \overline {\cal K}_{j}   \right) 
 \propto  S_{{\cal K}}  e^{-S_0}  $ where 
 $S_0 =\frac{4 \pi}{g^2 \beta_0}$. Note the appearance of the $\beta_0$ factor in the exponent, associated with the fractionalization of the  kink-saddles.

More interesting effects arise at second order in a semi-classical expansion. There are both  charged and neutral bion events contributing to the condensate, ${\cal B}_{ij} = [ {\K}_i \bar\K_j]  $,  
${\cal B}_{ii, \pm} = [\K_i\bar\K_i]_{\pm} $. The action of both events is $2S_0=\frac{8 \pi}{g^2 \beta_0}$. The crucial point  is that the neutral bion event is two-fold ambiguous, and this ambiguity is associated with the leading order growth (and hence ambiguity) of the perturbation theory.   
 \begin{eqnarray}
t_{\rm IR}^{\R \times S^1_L}=2 \times \frac{4 \pi}{\beta_0\, g^2(LN)} \leftrightarrow    {\cal B}_{ii, \pm} \sim e^{-\frac{8 \pi}{g^2 \beta_0}  } \pm i \pi  e^{-
\frac{8 \pi}{g^2 \beta_0} }  \sim  (\Lambda L)^4 \pm i  (\Lambda L)^4 
\label{two}
\end{eqnarray}
Studying perturbation theory at small circle in the effective dimensionally reduced QM system  produces an ambiguity exactly the same as in \eqref{two}, but opposite in sign. This is exactly the same effect as was observed in the $\mathbb {CP}^{N-1} $ model \cite{Dunne:2012ae}. 
Indeed, we see that the
 ambiguity in the spin-wave condensate calculated on  ${\R \times S^1_L}$    is sourced by neutral bions.   
 \begin{align}
 \label{ope-semi}
{\rm Im}  \langle \partial_{\mu} n^a  \partial_{\mu} n^a  \rangle_{\pm}   \propto {\rm Im}   {\cal B}_{ii, \pm} =  {\rm Im}  [\K_i\bar\K_i]_{\pm} 
 \end{align} 
 In the small-$L$ regime, the Borel plane singularities for the $O(N \geq 4)$ models is diluted by a factor of two, with respect to the singularities on $\R^2$ (similar to PCM model \cite{Cherman:2013yfa})
  while for $O(3)$ model, 
 the location of singularities remain unchanged. It remains an open question to understand the flow of singularity location as a function of compactification radius. 

\subsection{$O(3)$ model with  $\Theta$-angle}
\label{sec:O3theta}
The $O(3)$ model, unlike the $O(N \geq 4)$ model, admits a topological theta angle, and relatedly, instantons. In the $O(3)$ model,   there are two minimal action fractionalized
kink-instantons, each with topological charge $Q= \half$ \cite{Bruckmann:2007zh, Eto:2006mz}.   Since the $\Theta$ angle is periodic by $2\pi$,  the kink-instanton amplitudes  are  multi-branched, two-branched in this case.    The   amplitudes associated with these events are: 
 \begin{align}
& {\cal K}_{1,k}    = e^{-S_I/2} e^{ i \frac{\Theta + 2 \pi k}{ 2} }, \qquad     \overline {\cal K}_{1,k} = e^{-S_I/2} e^{ -i \frac{\Theta + 2 \pi k}{ 2} },  \qquad \cr
&  {\cal K}_{2,k}  = e^{-S_I/2} e^{ i  \frac{\Theta + 2 \pi k}{ 2} }, \qquad  \overline {\cal K}_{2,k}= e^{-S_I/2} e^{ -i  \frac{\Theta + 2 \pi k}{ 2} }.
 \end{align}
where $k=0,1$.  
Under a $2 \pi$ shift of the $\Theta$ angle, the  kink amplitude transforms as  
 \begin{align}
 \Theta \rightarrow \Theta + 2\pi:    \qquad & {\cal K}_{a,k} \rightarrow   {\cal K}_{a,k+1} 
 \end{align}
  reflecting the two-branched structure: each kink-instanton amplitude returns to itself under a $4\pi$ shift.  
On the other hand, the full 2d  instanton may be viewed as a composite of the these kink-instantons, and as expected,  it is independent of branch, manifestly periodic by $2 \pi$. 
Under a $2 \pi$ shift of the $\Theta$ angle, the  instanton amplitude is invariant:
 \begin{align}
 \Theta \rightarrow \Theta + 2\pi:   
\qquad {\cal I} \sim      {\cal K}_{1,k}  {\cal K}_{2,k}     
\sim  e^{-S_I} e^{ i \Theta } 
 \rightarrow {\cal K}_{1,k+1}  {\cal K}_{2,k+1} \sim {\cal I}
 \end{align}

The non-perturbatively induced mass gap of the theory is the non-perturbative energy splitting between the ground state and the first excited state. This is an effect which is induced by 
kink-instantons at leading order, similar to the discussion  for the $\mathbb {CP}^1$ in \cite{Dunne:2012ae}. 
The $\Theta$ angle  dependence of the mass gap at leading order in semi-classics is given by a two-branched function: 
 \begin{align}   
 \label{gap} 
m_g (\Theta)   
 =   {\rm Max}_k  \left[ \Lambda    \cos{ \textstyle{ \frac{\Theta 
 + 2 \pi k}{2} } }    +O(e^{-2S_0} ) \right] 
 \end{align}
For $\Theta=0$, mass gap is maximal. For $\Theta=\pi$, mass gap vanishes at leading order in semi-classics.  At sub-leading order, an exponentially smaller mass gap may be induced, this is sub-leading compared to the effects considered here.
This result of semi-classics provides evidence in favor of Haldane's conjecture,  identifying  $\Theta= 2 \pi S$,  where $S$ is spin.  Indeed, Haldane claimed that the integer spin theory  ($\Theta=0$)  is gapped, while the  half-integer spin theory   ($\Theta= \pi $), is gapless \cite{Haldane:1983ru}. Our leading semi-classical result is in concordance with this. 

 \begin{figure}[ht]
\begin{center}
\includegraphics[angle=0, width=3.in]{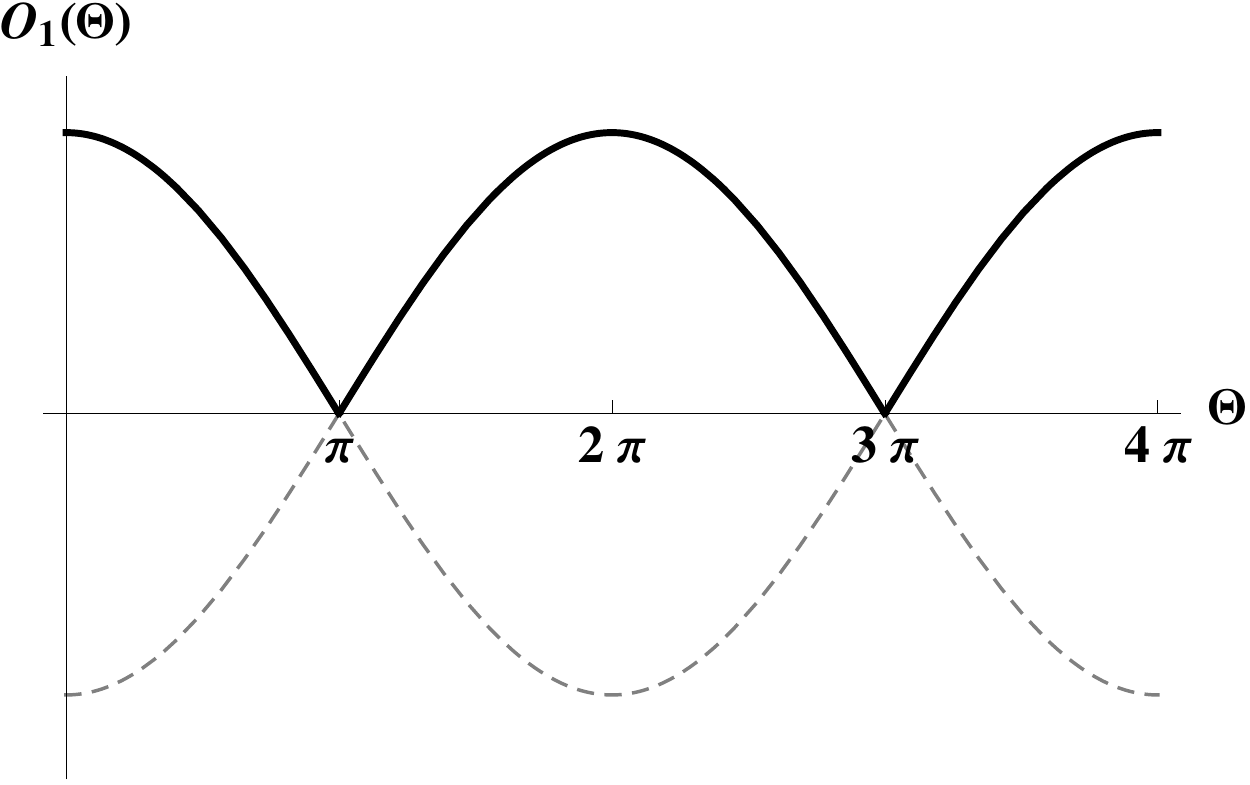}
\includegraphics[angle=0, width=3.in]{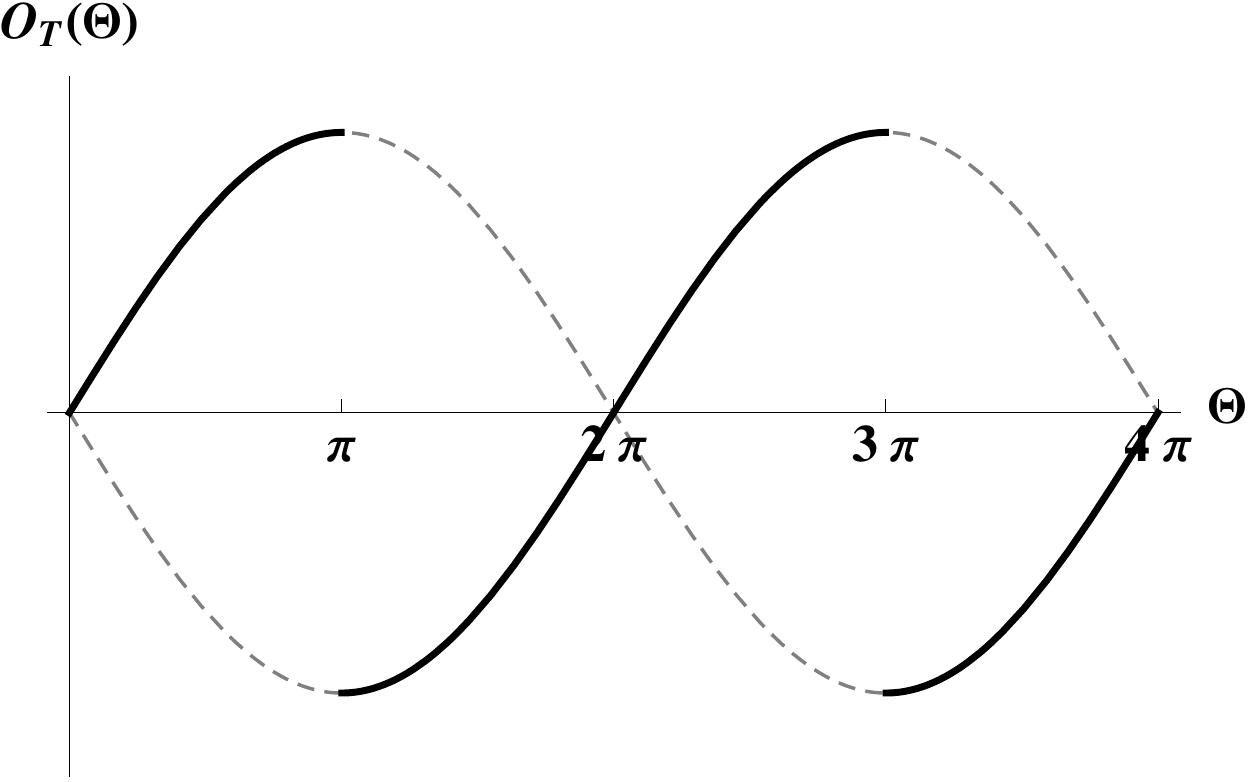}
\caption{The $\Theta$ angle dependence of  various observables: {\bf Left:} spin wave condensate $O_1(\Theta)$  and mass gap $m_g (\Theta)$. 
{\bf Right:}  Topological chage density condensate $O_T(\Theta)$
  in $O(3)$ model. Each is a two-branched function.  $O_1(\Theta)$ has a cusp at $\Theta=\pi$ associated with a change of branch. $O_T(\Theta)$ has a discontinuity at $\Theta=\pi$.  }
\label {fig:spinwave}
\end{center}
\end{figure}

As already mentioned for general $O(N)$, in the  Euclidean path integral representation. the  condensate  $O_1(\Theta) = \langle  \half \partial_{\mu} n^a  \partial_{\mu} n^a \rangle $ receives the leading non-perturbative contribution from kink-instantons.   Fig.~\ref{fig:dkbg} represents a snap-shot of the Euclidean vacuum of the theory. It is, as described already, a dilute gas of one and two-events. Calculating the  vacuum expectation value $O_1(\Theta) $  at $\Theta=0$ and at leading order in semi-classics   is equivalent to finding the average of the  action density  $ \half \partial_{\mu} n^a  \partial_{\mu} n^a $ over all space.  Introducing  $\Theta \neq 0$ is equivalent to  a complex phase for the fugacity of topological defects, a complex fugacity. 
  The action density  is concentrated  within the characteristic size $r_K $  of the kink-instanton.

The kink-instantons contribute in the same way to $O_1(\Theta) $ and   topological charge density condensate  $O_T(\Theta)  =  \langle  \frac{1}{8\pi} \epsilon_{\mu \nu} \epsilon_{abc} 
{ n_a}  \partial_\mu  { n_b}   \partial_\nu  {n_c} \rangle $   as    $+S_{\cal K} {\cal K}_{a,k} $, because both configuration has the same action density. 
 On the other hand, 
the anti-kink-instantons contribute to  $O_1(\Theta) $  as $ +S_{\cal K} \overline {\cal K}_{a,k} $ while they contribute to $O_T(\Theta) $ as 
 $ - S_{\cal K} \overline {\cal K}_{a,k} $, because the topological charge densities of the two configurations are opposite in sign.   
   Therefore, the leading contribution to these two condensates takes the form:  
\begin{align}
 O_1(\Theta)    \propto   +S_{\cal K}  {\rm Ext}_k  \left( {\cal K}_{a,k}   + \overline {\cal K}_{a,k}   \right) 
 \propto  S_{\cal K} e^{-S_I/2}   {\rm Ext}_k    \cos{  \frac{\Theta + 2 \pi k}{ 2} }  \cr
 O_T(\Theta)    \propto  +S_{\cal K}   {\rm Ext}_k  \left(  {\cal K}_{a,k}   -   \overline {\cal K}_{a,k}  \right)  \propto
 S_{\cal K} e^{-S_I/2}   {\rm Ext}_k    \sin{  \frac{\Theta + 2 \pi k}{ 2} }  
 \label{two-condensate}
\end{align} 
where by ${\rm Ext}_k$, we mean that the vacuum energy should be extremized among the two-branches, and the observable should always be calculated at the genuine vacuum branch.    

 The $O(3)$-model Lagrangian has an exact CP-symmetry at $\Theta=0$ and at  $\Theta=\pi$.  Under CP-symmetry, $i \Theta\, Q_T \rightarrow - i \Theta\, Q_T$, and 
this is a symmetry if and only if $\Theta=0, \pi$, because  $\Theta$ is periodic by $2\pi$.  

At  $\Theta=0$,   the $O_1(\Theta)$ condensate is maximal, and  the $O_T(\Theta)$ condensate vanishes.  The reason for the vanishing of the latter is that 
kinks and anti-kinks contribute oppositely to the topological charge density condensate. A consequence of this is  unbroken CP-symmetry.

At  $\Theta=\pi$,   the $O_1(\Theta)$ condensate vanishes, due to topological interference, and  the $O_T(\Theta)$ condensate is discontinuous, i.e., it jumps.  This discontinuity and jump are not related to resurgence.   Instead,   $O_T$ is an order parameter for CP-symmetry, and this symmetry is believed to be spontaneously broken at $\Theta=\pi$ on $\R^2$, taking one of the two possible values:   $\langle O_T \rangle = \pm \Lambda^2 $, and two isolated vacua.  Our leading semi-classical analysis confirms this expectation.

\section{Grassmannian model}
 The Grassmannian model, denoted as ${\rm Gr}(N, M)$, is a 2d non-linear sigma model with complex Grassmannian target space:
 \begin{eqnarray}
{ \rm Gr}(N, M)=  \frac{U(N)}{U(N-M) \times U(M)} 
 \label{eq:grassmannian}
 \end{eqnarray}
These models have instantons  \cite{zak-book,Macfarlane:1979hi,Din:1981bx}.
For $M=1$, ${\rm Gr}(N, M)$ reduces to the $\mathbb {CP}^{N-1}$ model whose non-perturbative resurgent properties were studied in  \cite{Dunne:2012ae}.   There a key feature of the analysis was the fractionalization of $\mathbb {CP}^{N-1}$ instantons into $N$ fundamental kink-instanton components when compactifed with twisted boundary conditions, and the correspondence with the beta function coefficient $\beta_0=N$ for $\mathbb {CP}^{N-1}$. Here we show that the situation for the Grassmannian models ${\rm Gr}(N, M)$ with $M\geq 2$ is actually quite similar. 
Some aspects of classifying  bion solutions in Grassmannian models have also appeared in \cite{Misumi:2014bsa}, but here we study the fractionalizations associated with adiabatic spatial compactification. The beta function coefficient $\beta_0=N$, independent of $M$, and the twisting is again associated with the global $U(N)$ symmetry group. Moreover, the construction of classical solutions is quite  similar for all ${\rm Gr}(N, M)$.  But there are some interesting technical differences, which we outline in this section.
    For example, it is not immediately clear 
 what is the most useful parametrization of the ${\rm Gr}(N, M)$ manifold such that we can see the fractionalization of 2d instanton into kink-instantons in its simplest form. 
 It turns out that once this is achieved, 
 the construction of two-events (neutral and charged bions) and the relation of these semi-classical configurations to renormalons follows  the universal pattern of gauge theory and other non-linear sigma models, 
such as   $\mathbb {CP}^{N-1}$.  

For  ease of presentation, we  first introduce some notation. 
We define the theory  on a two-dimenisonal manifold,  the plane  $M_2= \R^2$, and the spatially compactified cylinder $M_2=  \R \times  S^1_L$.  Our methods easily generalize 
to theories  with $N_f$ fermionic species, similar to \cite{Dunne:2012ae}.    The bosonic field  is defined as the map:
\begin{eqnarray}
 z(x)\;  : \; M_2 \rightarrow  {\rm Gr}(N, M)\equiv  \frac{U(N)}{U(N-M) \times U(M)} 
\label{action-gr}
\end{eqnarray} 
The real dimension of the  ${\rm Gr}(N, M)$ manifold is equal to the number of microscopic independent degrees of freedom in the model:
 \begin{eqnarray}
{\rm dim}_{\R}{\rm Gr}(N, M)= N^2-[ (N-M)^2 +M^2] = 2M(N-M)
\label{dim-gr}
\end{eqnarray} 
This aspect is interesting, because as $M$ interpolates from $O(N^0)$ to $O(N^1)$, the number of degrees of freedom interpolates from being $O(N^1)$ to $O(N^2)$.  The theory moves from a vector-like   to a matrix-like large-$N$ limit.
Explicitly,  $z(x) $ is an $N \times M$  matrix 
\begin{align}
(z)_{ja}(x), \qquad j=1, \ldots, N, \qquad a=1, \ldots M
 \end{align}
 obeying the  constraint: 
 \begin{align}
z^{\dagger}z={\bf 1}_{M\times M}
\label{constraint}
\end{align}
The  action of the bosonic model is 
\begin{eqnarray}
S=\frac{2}{g^2}\int_{M_2}  \widetilde \tr  \left |D_\mu z\right|^2  - i   \frac{\Theta}{2\pi}  \int_{M_2}   \epsilon_{\mu \nu}  \widetilde \tr \partial_{\mu} z^{\dagger} \partial_\nu z  \quad, \quad D_{\mu}z = \partial_{\mu} z - z A_{\mu}
\label{action-gr-0}
\end{eqnarray} 
where $A_{\mu}=z^{\dagger} \partial_{\mu} z$  is an $M\times M$ auxiliary gauge field,   $D_{\mu}z = \partial_{\mu} z - z A_{\mu}$ is gauge covariant derivative, and $\Theta$ is topological theta-angle.  The traces   is over the space of $M\times M$ matrices. 
 with the convention  $ \widetilde \tr  = \frac{1}{M} \tr$. 
  The action (\ref{action-gr-0}) has a  global $U(N)$ symmetry, and a  local $U(M)$ gauge redundancy   
  under which the elementary field transforms as:
\begin{eqnarray}
z(x)\to \Omega_N   z(x) \Omega_M(x)  \quad,\quad \Omega_N \in U(N), \qquad \Omega_M \in U(M),
\label{global}
\end{eqnarray}
We   will use  \eqref{global} to impose twisted boundary conditions on the cylinder $\R \times S^1_L$. 

The $z(x)$-field is massless classically, and also  to all orders in perturbation theory. 
The bosonic  theory on $M_2=\R^2$ is believed to possess a mass gap quantum mechanically, although the mechanism  via which a mass gap is formed is unknown on $\R^2$.   
The theory possess 2d instantons, solutions to the self-duality equations, 
$D_\mu z= \pm i \epsilon_{\mu \nu} D_\nu z$, with quantized action and relation to strong scale:
\begin{align}
S_{\cal I} &= \frac{  4 \pi }{g^2}, \qquad {\cal I}_{2d} \sim e^{- \frac{  4 \pi }{g^2(\mu)}},   
\qquad  \Lambda^{ \beta_0} = \mu^{ \beta_0} {\cal I}_{2d}, \qquad {\rm any} \; (N, M)
\end{align}
The leading perturbative beta function coefficients is
\begin{eqnarray}
\beta_0 =N\quad \text{for}\quad  {\rm Gr}(N, M)
\end{eqnarray}
independent of $M$.   It is also unchanged by the inclusion of $N_f$ fermion flavors. The theory is asymptotically free for all $(N, M)$ and for all $N_f$.

Here, we study this ${\rm Gr}(N, M)$ theory in a  semi-classically calculable regime, on $\R \times S^1_L$ with twisted boundary conditions:
\begin{eqnarray}
z(x_1, x_2+L) &=& \Omega_N^0  z (x_1, x_2)  \Omega_M^{0} \cr
z_{ja}(x_1, x_2+L) &=& e^{2 \pi i \mu_j}  z_{ja}(x_1, x_2)  e^{-2 \pi i \omega_a}  
\label{tbc-3}
\end{eqnarray}
The  choice of $\omega_a$ is irrelevant, 
because the $U(M)$ ``symmetry''  is actually a (gauge) redundancy, while the $U(N)$ symmetry is a global one. Explicit calculation also shows that this is indeed the case. 

\subsection{Explicit construction for ${\rm Gr}(N, 2)$}
Below, we give details of the explicit construction of kink-instanton solutions for  ${\rm Gr}(N, 2)$, the first class of non-$\mathbb {CP}^{N-1}$ Grassmannian theories. In order to find the counter-part of the kink-instantons in  $\mathbb {CP}^{N-1}$, 
we first project to subspaces in which the  minimal tunneling events take place:
\begin{equation} 
\label{grn2}
z= \begin{pmatrix}
z_{11} &z_{12} \cr
z_{21} & z_{22}\cr
\vdots & \vdots \cr
\vdots & \vdots \cr
z_{N1} & z_{N2} \cr
\end{pmatrix} \longrightarrow z= \begin{pmatrix}
0 &0  \cr
\vdots & \vdots \cr
z_{j1} & z_{j2}\cr
z_{j+1,1} & z_{j+1,2}\cr
\vdots & \vdots \cr
0&0  \cr
\end{pmatrix} 
\end{equation} 
The first column of \eqref{grn2} can be picked as a representation of  $\mathbb {CP}^{N-1}$. Once this is done, the constraint \eqref{constraint} restricts the form of the second column:  
\begin{equation} 
 z=\begin{pmatrix}
0 &0  \cr
\vdots & \vdots \cr
e^{i \phi_{j1}} \cos \frac{\theta}{2}  & - e^{i \phi_{j2} } \sin \frac{\theta}{2}   \cr
 e^{i \phi_{j+1,1}} \sin \frac{\theta}{2}  &   e^{i \phi_{j+1,2}} \cos \frac{\theta}{2}   \cr
\vdots & \vdots \cr
0&0  \cr
\end{pmatrix}  \qquad \text{such that},  \;
\; \phi_{j1}- \phi_{j+1, 1} =  \; \phi_{j2}- \phi_{j+1, 2} \; . 
\end{equation} 
In this parametrization, twisted boundary conditions act only on $ \arg z_{ja} =\phi_{j,a}  $. 
\begin{align}
\phi_{ja}  (x_1, x_2+L)  = \phi_{ja}  (x_1, x_2) + 2 \pi \mu_j - 2 \pi \omega_a 
 \end{align} 
The Lagrangian is a trace over a $2 \times 2$ matrix, whose diagonals are  ${\cal L}_{aa}$, $a=1,2$
\begin{align}
 {\cal L}_{aa}=  \frac{1}{2 g^2}  \left[ (\partial_\mu \theta)^2 + 
\sin^2 \theta   (\partial_\mu (\phi_{ja} -\phi_{j+1,a}) + \xi \delta_{\mu2})^2   \right], \qquad \xi= 2 \pi  (\mu_j -   \mu_{j+1}) 
\label{action-gr}
\end{align} 
where the    twisted boundary conditions are undone in favor of a  background  field.  Note that in the combination 
$ (\phi_{ja} -\phi_{j+1,a})$, the dependence on $\omega_a$ drops out. This  is because the $U(2)$  ``symmetry"  is a local redundancy, as opposed to being a genuine symmetry.
Thus,  ${\cal L}_{11} =  {\cal L}_{22}$. 

\subsection{One- and two-events: Kink-instantons and bions}
The action \eqref{action-gr}  is essentially the same as what appeared in the $\mathbb {CP}^1$ model.  When reduced to quantum mechanics, 
we obtain the same quantum mechanical action \eqref{actionQM}.  
It has obvious  minimal action kink-instanton solutions associated with the simple roots of the $\frak u(N)$ algebra. There is also a kink-instanton 
event associated with the affine root, whose demonstration is identical to the discussion of $\mathbb {CP}^1$.

Kink-instanton  events interpolate between the minima of the action and their explicit form is given by 
\begin{align} 
{\cal K}_1:& \begin{pmatrix}
1 &0 \cr
0 & 1 \cr
0 & 0\cr
\vdots & \vdots \cr
0 & 0\cr
\end{pmatrix} \longrightarrow 
\begin{pmatrix}
0 &-1\cr
1 & 0 \cr
0 & 0\cr
\vdots & \vdots \cr
0 & 0\cr
\end{pmatrix} \qquad 
{\cal K}_2: \begin{pmatrix}
0 &0 \cr
1 & 0  \cr
0 & 1 \cr
\vdots & \vdots \cr
0 & 0\cr
\end{pmatrix} \longrightarrow 
\begin{pmatrix}
0 & 0\cr
0 & \textstyle{ -1} \cr
1 & 0 \cr
\vdots & \vdots \cr
0 & 0\cr
\end{pmatrix} \qquad,  \ldots,  \cr
{\cal K}_{N-1}:& \begin{pmatrix}
0 &0 \cr
\vdots & \vdots \cr
0 &0 \cr
1 & 0 \cr
0 & 1\cr
\end{pmatrix} \longrightarrow 
\begin{pmatrix}
0 & 0\cr
\vdots & \vdots \cr
0 & 0\cr
0 & -1\cr
1 & 0
\end{pmatrix} \qquad 
{\cal K}_N: \begin{pmatrix}
0 &1 \cr
0 & 0  \cr
\vdots & \vdots \cr
0 & 0\cr
1 & 0\cr
\end{pmatrix} \longrightarrow 
\begin{pmatrix}
1 &0 \cr
0 & 0  \cr
\vdots & \vdots \cr
0 & 0\cr
0 & -1\cr
\end{pmatrix}
\end{align} 
In the space of fields, these tunneling events correspond to the change in the $z$ field by an amount:
\begin{align}
{\cal K}_j : e_j + f_{j+1} \rightarrow e_{j+1} - f_{j} \qquad \Delta z = z(\infty)- z(-\infty) \equiv   \alpha_j =  e_{j+1} - e_j - f_{j+1}  - f_{j} 
\end{align}
The non-perturbative weight associated with these kink-instanton saddles is given by 
\begin{align}
{\cal K}_j \sim e^{-S_{\cal I}/N} = e^{- \frac{  4 \pi }{g^2N}},   \qquad   j=0,1, \ldots, N-1
\end{align}
The connection between these twisted semi-classical solutions and the IR renormalons in perturbation theory then follows exactly the same construction as for the $\C\P^{N-1}$ model \cite{Dunne:2012ae}.


\subsection{$\Theta$-angle dependence in ${\rm Gr}(N, 2)$}
${\rm Gr}(N, 2)$ model  admits the addition of a topological theta-term to the microscopic Lagrangian, $ i \Theta\,  Q_T$, written explicitly in 
\eqref{action-gr-0}.
For a  2d instanton, the topological charge is $Q_T=1$.  
As discussed above, there are $N$ minimal action 
kink-instantons, each with topological charge $Q_T= \frac{1}{N}$.  Since $\Theta$ angle is periodic by $2\pi$,  the kink-instanton amplitudes  are  multi-branched (here $N$-branched), just 
as for the monopole-instanton amplitudes in   (deformed) Yang-Mills theory on ${\mathbb R}^3 \times S^1_L$ \cite{Unsal:2012zj,Bhoonah:2014gpa}.  
   The   amplitudes  associated with such  kink-instanton or kink-anti-instanton events are: 
 \begin{align}
& {\cal K}_{j,k}    = e^{-S_I/N} e^{ i \frac{\Theta + 2 \pi k}{ N} }, \qquad     \overline {\cal K}_{j,k} = e^{-S_I/N} e^{ -i \frac{\Theta + 2 \pi k}{ N} },   \qquad \; k=1, \ldots, N
 \end{align}
 for each $j=0, 1, \ldots, N-1$.
  Under a $2 \pi$ shift of the $\Theta$ angle, each  kink-instanton  amplitude  transforms cyclically: 
 \begin{align}
 \Theta \rightarrow \Theta + 2\pi:    \qquad & {\cal K}_{j,k} \rightarrow   {\cal K}_{j,k+1}   
 \end{align}
  reflecting  the $N$-branched structure.  ${\cal K}_{j,k}$   returns to itself after   $2\pi N$ shifts in $\Theta$,   ${\cal K}_{j,N+k} = {\cal K}_{j,k}$.  
The 2d  instanton is a composite of the $N$-kink instantons, and as expected,   it is independent of branch: 
 \begin{align}
 {\cal I} \sim   \prod_{j=0}^{N-1}   [  {\cal K}_{j,k} ]^{k_j^\v} =\prod_{j=0}^{N-1}    {\cal K}_{j,k} 
   =  e^{-S_I} e^{ i \Theta } = e^{ -\frac{4 \pi}{g^2} + i \Theta}
  \end{align}
  Under a $2 \pi$ shift of the $\Theta$ angle, the instanton amplitude is invariant:  ${\cal I}         \rightarrow  {\cal I}$.

 \begin{figure}[ht]
\begin{center}
\includegraphics[angle=0, width=3.in]{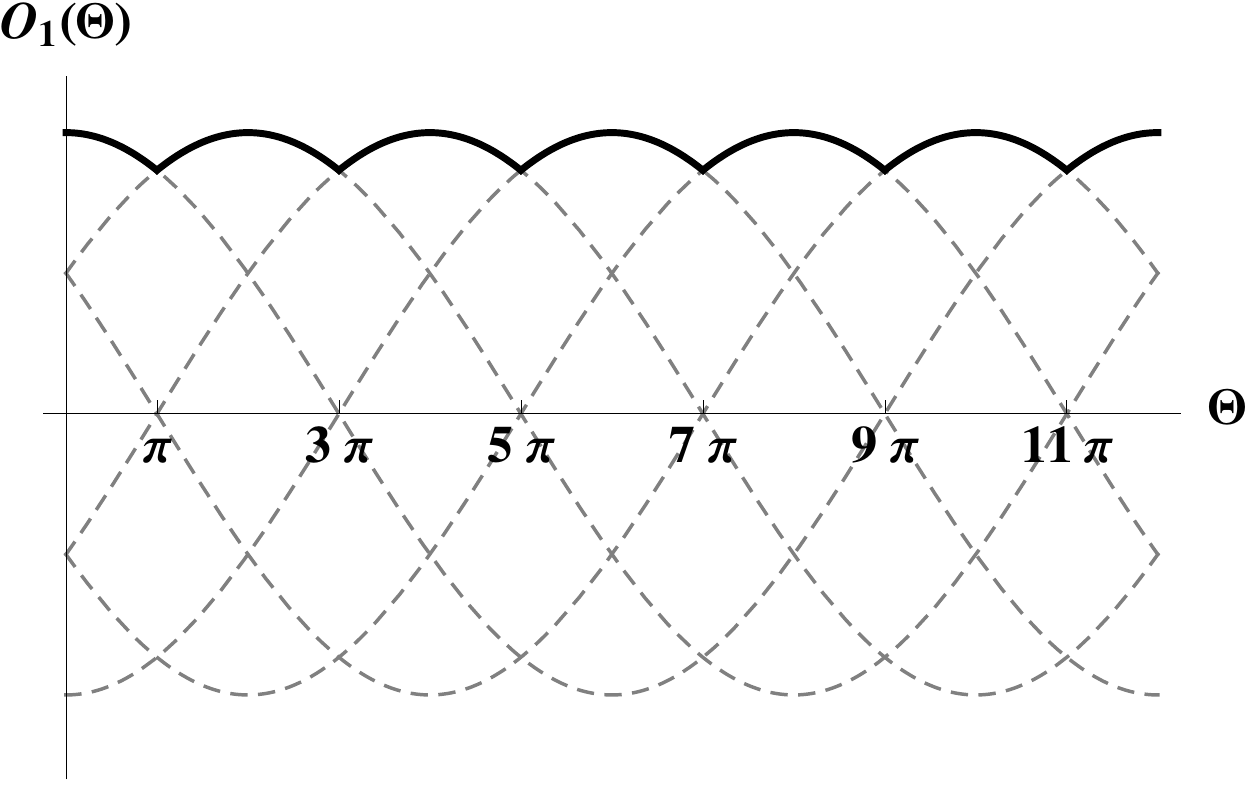}
\includegraphics[angle=0, width=3.in]{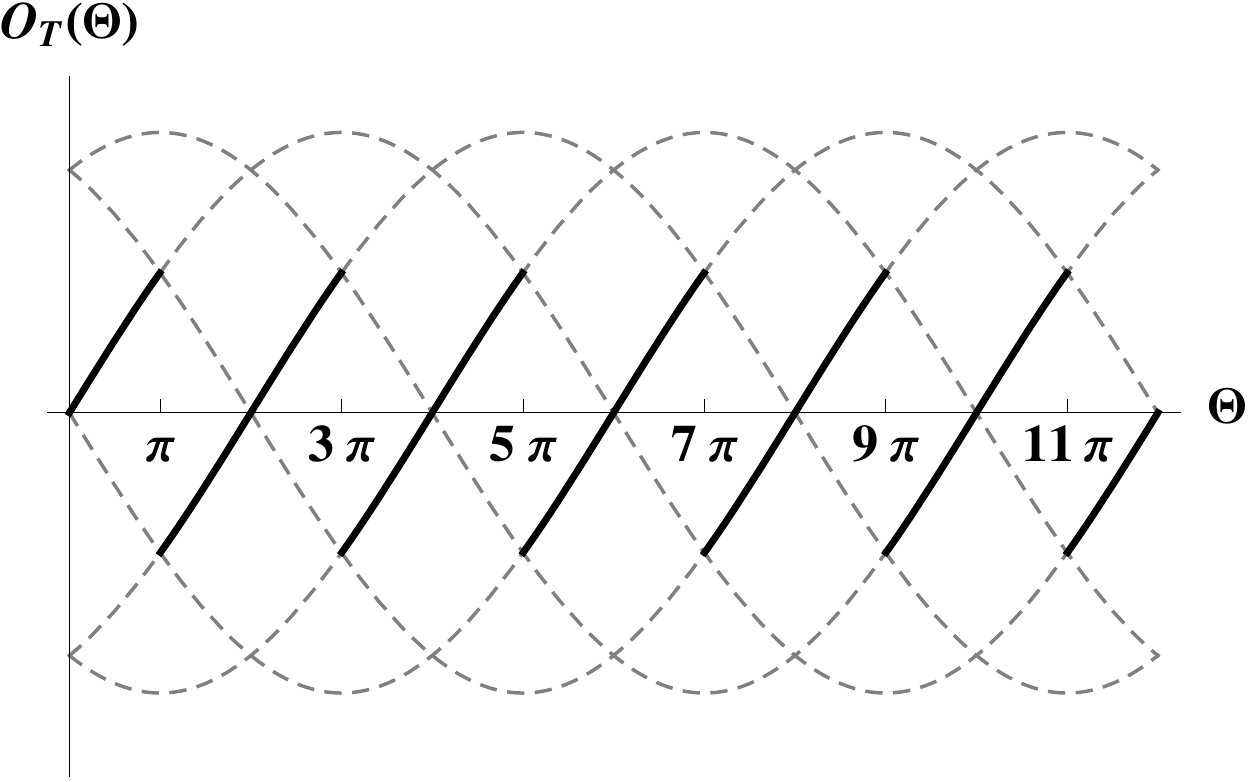}
\includegraphics[angle=0, width=3.in]{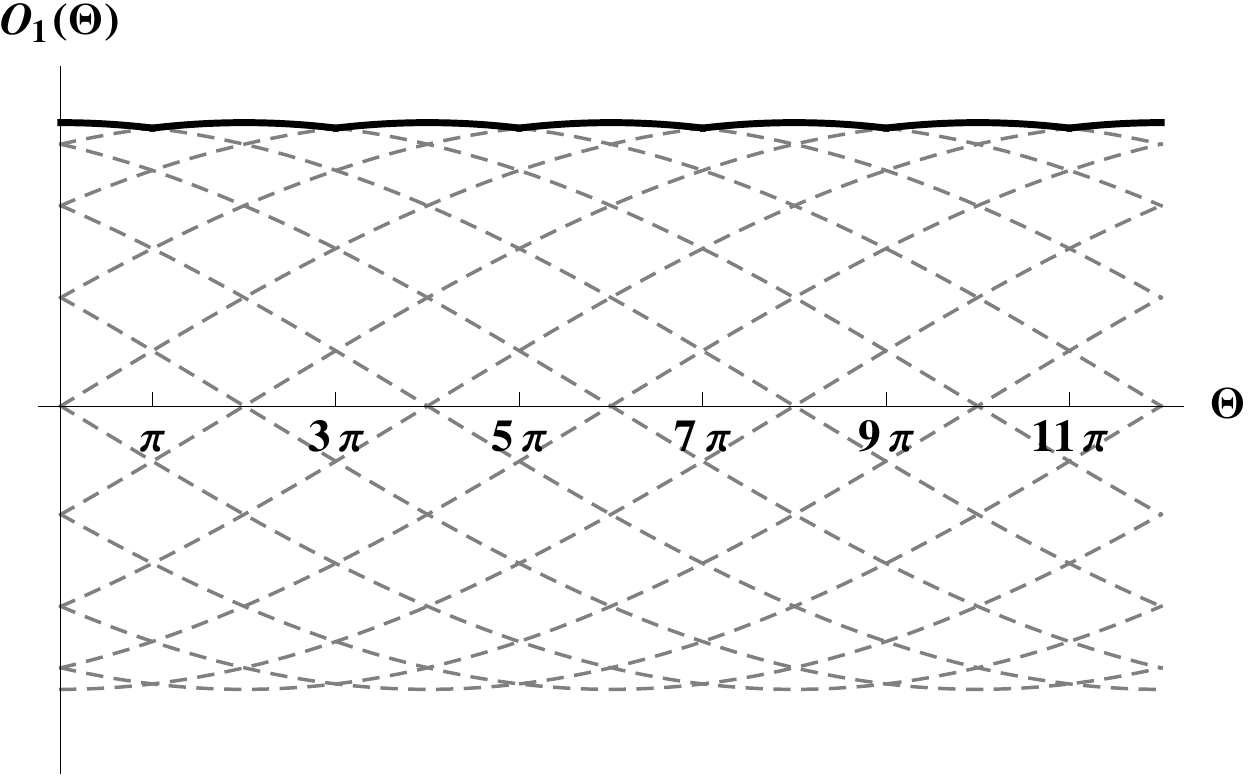}
\includegraphics[angle=0, width=3.in]{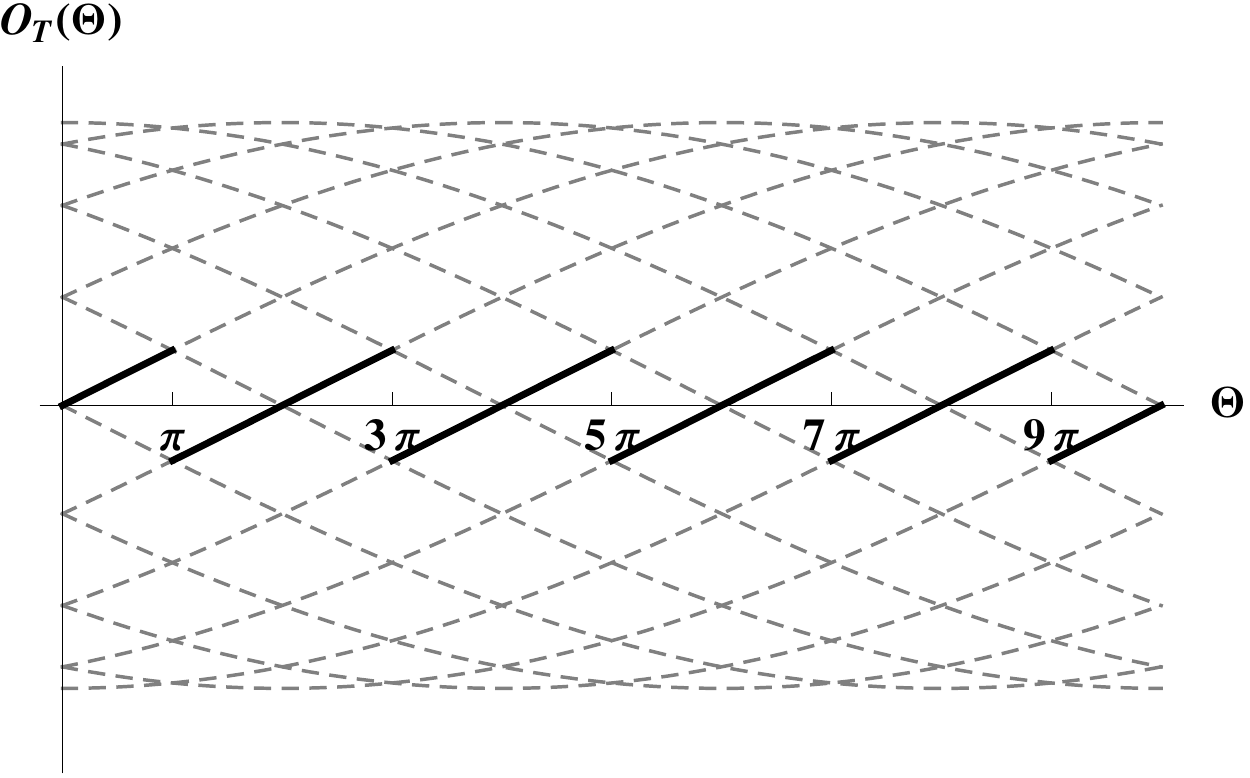}
\caption{The $\Theta$ angle dependence of   condensate $O_1(\Theta) $, and topological chage density condensate $O_T(\Theta)$ in ${\rm Gr}(N,2)$ model for $N=6$ (top) and $N=16$ (bottom). Both are $N$-branched function.  $O_1(\Theta)$ has a cusp at $\Theta=\pi$ associated with a change of branch. $O_T(\Theta)$ has a discontinuity at $\Theta=\pi$.  }
\label {fig:condensate-Gr}
\end{center}
\end{figure}

 The discussion of the condensates follows almost verbatim that of the $O(3)$ model in Section \ref{sec:O3theta}.  
Consider the  operators, 
\begin{align}
     {O}_1 ={\cal L} ,  \qquad         { O}_T= Q_T,       
 \end{align} 
 and  their vacuum expectation values, i.e., the condensates.  
The leading semi-classical contribution to these two condensates takes the form:  
\begin{align}
 O_1(\Theta)    \propto   +S_{\cal K}  {\rm Ext}_k  \left( {\cal K}_{j,k}   + \overline {\cal K}_{j,k}   \right) 
 \propto  S_{\cal K} e^{-S_I/N}   {\rm Ext}_k    \cos{  \frac{\Theta + 2 \pi k}{ N} }  \cr
 O_T(\Theta)    \propto  +S_{\cal K}   {\rm Ext}_k  \left(  {\cal K}_{j,k}   -   \overline {\cal K}_{j,k}  \right)  \propto
 S_{\cal K} e^{-S_I/N}   {\rm Ext}_k    \sin{  \frac{\Theta + 2 \pi k}{ N} }  
 \label{two-condensate-Gr}
\end{align} 
where by ${\rm Ext}_k$, we mean that the vacuum energy should be extremized among $N$ branches, and the observable should always be calculated at the genuine vacuum branch.  For generic $\Theta$,  the vacuum  branch is unique, and only  for  $\Theta=\pi$, it is two-fold degenerate, where there exist two vacua as described below. 
The condensates  are plotted in  Fig.~\ref{fig:condensate-Gr}. 

 The ${\rm Gr}(N, 2)$-model Lagrangian has an exact CP-symmetry at $\Theta=0$ and at  $\Theta=\pi$.  Under CP-symmetry, $i \Theta\, Q_T \rightarrow - i \Theta\, Q_T$, and 
this is a symmetry if and only if $\Theta=0, \pi$, because  $\Theta$ is periodic by $2\pi$.   This symmetry is  believed to be spontaneously broken at 
$\Theta=\pi$ on $\R^2$.  

 In the semi-classical domain,  the highest deviation of the physical observables from their value at $\Theta=0$ happens at $\Theta=\pi$.  
\begin{align}
O_1(\Theta=\pi)  &\propto  \Lambda^2   (\Lambda LN)^{-1} \cos{    \frac{ \pi }{ N} }  \underbrace{\longrightarrow}_{ \Lambda LN \sim 1} 
       \Lambda^2 \cos{    \frac{ \pi }{ N} }   \cr
    { O}_T (\Theta=\pi)   & \propto   \pm  \Lambda^2   (\Lambda LN)^{-1}\sin{    \frac{ \pi }{ N} }  \underbrace{\longrightarrow}_{ \Lambda LN \sim 1} 
    \pm   \Lambda^2 \sin{    \frac{ \pi }{ N} } 
    \end{align}  
This is an interesting result for two different reasons, one related to CP-realization, and the other large-$N$ dynamics. 
 
At  $\Theta=\pi$,   the $O_1(\Theta)$ condensate has a cusp, and reaches to its minimum, while the $O_T(\Theta)$ condensate is discontinuous. As in $O(3)$ case,  this discontinuity is not related to resurgence.   Instead,   $O_T$ is an order parameter for CP-symmetry, and this symmetry is believed to be spontaneously broken at $\Theta=\pi$ on $\R^2$, taking one of the two possible values:   $\langle O_T \rangle  \propto  \pm \Lambda^2 $.  Our leading semi-classical analysis confirms this expectation. 
In quantum mechanics, these  two sectors are similar to different superselection sectors.

  At the scale where the weak-coupling approximation breaks-down, $\Lambda LN \sim 1$,  both the $O_1$ and $O_T$  condensates reach to their   ``natural"    $\Lambda^2$ scaling.  However,  the maximum (in magnitude)  value that  $O_T$ can take while staying on the vacuum branch is actually suppressed by a factor  of $ \sin{    \frac{ \pi }{ N} } \sim \frac{1}{N}$.    On the other hand, for  $O_1 (\Theta)$,    the maximal deviation  (with respect to $\Theta=0$)  that it exhibit  while  staying on the vacuum branch is actually suppressed by a factor  of $  1/N^2$, see  Fig.~\ref{fig:condensate-Gr} for N=6 and 16, for example. 
  
 In the $N=\infty$ limit, all (non-extensive) observables with $O(N^0)$ scaling must be independent of the $\Theta$-angle, as shown in \cite{Unsal:2012zj}.  Indeed, using \eqref{two-condensate-Gr}
  we observe that 
  \begin{align}
 O_1(\Theta)   = c_1  \Lambda^2 , \qquad   O_T(\Theta) =0  \qquad {\rm at}  \; N=\infty.
  \label{two-condensate-Gr-2}
\end{align} 
This is an implication of  large-$N$ theta-angle independence.  
This also means that  spontaneous CP-breaking  does not appear at leading large-$N$ dynamics, and only appears at $1/N$ level.  This anticipation based on the semi-classics can be studied exactly by using exact large-$N$ solution of the model.

\section{Conclusions}
Our results for the $O(N)$ model compactified on $\R^1 \times S^1$ are  surprising, in part  because they show remarkable similarities with gauge theories. While the 2d $O(N)$ sigma models have been studied for  decades, it has generally been believed that the  analogy between these non-linear sigma models 
and four-dimensional gauge theories, in particular QCD,   breaks down once one considers  $O(N)$ models with $N \geq 4$,  because  these sigma models do not have instantons.  
Our results  indicate a different picture than this historical one. We have provided a physical interpretation of the 2d-saddles (non-instanton finite action classical solutions) in these sigma models.   We have shown that in a controlled weak-coupling semi-classical analysis of 2d sigma  models  on   $\R^1 \times S^1$, 
there exist kink-saddles in one-to-one correspondence with the affine root system of the ${\frak o}(N) $ algebra.  Furthermore, we have  shown the existence of a  resurgent structure in which classical  kink-anti-kink saddles produce non-perturbative contributions which cancel ambiguities arising from the Borel non-summability of  perturbation theory. 
This may be viewed as a semi-classical realization of the IR-renormalon, and may also provide a key part of the bridge to renormalons and the operator product expansion. 

Comparing with earlier work  in $SO(N)$ gauge theories \cite{Argyres:2012ka}, we also observe that  there exists a one-to-one mapping between  kink-saddles in the $O(N)$  sigma models  on  $\R^1 \times S^1$ and monopole-instantons in  $O(N)$ gauge theories on $\R^3 \times S^1$. This can be summarized in an elegant Lie algebraic relation 
between the 2d-saddle, and kink-saddle amplitudes, identical to the relation between 4d-instanton and monopole-instanton amplitudes:
    \begin{align}
     {\cal S}_{2d}  \sim  
 \prod_{j=0}^{\frak r} [{\cal K}_j]^{k_j^\v} \qquad   {\rm vs.}   \qquad   {\cal I}_{4d}  \sim  
 \prod_{j=0}^{\frak r} [{\cal M}_j]^{k_j^\v},  
  \qquad      {\rm where}  \;\;
h^\vee= \sum_{i=0}^{{\frak r}  } k_i^\vee.
  \end{align}
 It is natural to expect that the appearance of the dual Kac-labels as degeneracies of the kinks (or monopoles)  inside either the 2d-saddle (or 4d-instanton) is a universal phenomenon. 
In other words,   the 2d saddles fractionalize in the weak-coupling analysis of the twisted-compactified theory, in a manner directly related to the associated beta function, providing a key part of the bridge to renormalons and the operator product expansion.   This provides a new way to interpret the fact that the leading beta function coefficient is given generally by the dual Coxeter number: $\beta_0=h^\vee$.
 
 We have also analyzed the dependence on the topological theta angle in the $O(3)$ and ${ \rm Gr}(N, M)$ models,  explaining the multi-branched structure of observables in terms of the multi-branched structure of the kink-amplitudes. The results obtained via our formalism are consistent with what is known in the large $N$ limit. 
 
  Our analysis gives a concrete  physical interpretation of finite action saddle solutions, and motivates future work to  understand more systematically the fluctuation modes about non-BPS saddles in these sigma models and also in Yang-Mills theories where finite action non-BPS  saddles exist, but for which much less is known about their classification and fluctuations \cite{burzlaff,SSU,bor,Sadun:1992vj}.
  
 { \bf Cusp anomalous dimension in ${\cal N}=4$ SYM.  }
The $O(6)$ model where the field is valued on $S^5$, is also important in the context of gauge-string duality  
 \cite{Alday:2007mf}.
  It is known that the strong coupling expansion of the  cusp anomalous dimension in 
  ${\cal N}=4 $  supersymmetric Yang-Mills   is given by a Borel-non-summable asymptotic expansion  \cite{Basso:2007wd}.
Borel resummation of the cusp-anomalous dimension  is ambiguous with a jump given by  ${\cal S}_{\varphi=0^+} \Gamma_{\rm cusp}  - {\cal S}_{\varphi=0^-} \Gamma_{\rm cusp} \sim   i e^{\frac{- 2 \pi}{g^2} }\sim i m^2 $. In this work, we have given evidence that 
observables in the  $ O(6)$ theory are resurgent  and the ambiguities cancel, and in particular, the  neutral bion  $[{\cal B}_{ii}]_{\varphi=0^\pm}  $is two-fold ambiguous, and  cancels  the  ambiguity of perturbation theory in the semi-classical regime,
$\Im {\cal S}_{\varphi=0^\pm}  {\cal E}_{0}  + \Im\, [{\cal B}_{ii}]_{\varphi=0^\pm} =0 $.
In the strong coupling regime, this is presumably replaced by  
 $\Im {\cal S}_{\varphi=0^\pm}  {\cal E}_{0}  + {\rm Im}  \langle \partial_{\mu} n^a  \partial_{\mu} n^a  \rangle_{ \varphi=0^\pm} =0$,
  because of the relation \eqref{ope-semi} connecting the condensate appearing in the OPE to a semi-classical configuration.  
    In our context, it is clear in the semi-classical domain that the first order effect in the semi-classical expansion, 
namely  the kink-saddles,  leads to  the mass gap, and the second order effect in semi-classics lead to the cancellation of the ambiguity in the vacuum energy, and this is ultimately related to renormalons.  Our work suggests that the strong coupling expansion of the cusp-anomalous dimension is actually resurgent. 
It would be interesting to interpret  the renormalons or neutral bion from the string theory side.

{\bf  Fermions, hidden topological angle and vanishing condensates:} Although we did not discuss details of the inclusion of fermions, this can be done along the lines of our earlier work on 
${\mathbb CP}^{N-1}$ \cite{Dunne:2012ae}.   One effect that is not discussed in   \cite{Dunne:2012ae}  in the presence of fermions is the implication of hidden topological angles 
for the spin wave condensate.   Consider for example, the $N_f=1$ case, corresponding to the supersymmetric version of the $O(N)$ model or ${\rm Gr}(N,M)$ model. 
The spin-wave condensate is an order parameter for supersymmetry breaking, as can be deduced from the trace anomaly relation.  Since the Witten index of these  theories is, respectively,  
$I_W=2$, and $I_W=N$, supersymmetry is unbroken. As a result, the spin-wave condensate must vanish.  The microscopic mechanism for the vanishing of the condensate is actually 
 interesting and follows the same pattern as in the examples in \cite{Behtash:2015kna}. 
The kink-saddles do not contribute to the condensate because of the effect of  fermionic zero modes.  However, at second order in the semi-classical expansion, there are neutral bions and charged bions, which do contribute to the spin-wave condensate. As explained in   \cite{Dunne:2012ae},   ${\cal B}_{ii}$ is unambiguous  in the $N_f=1$ case, but there is a 
 a hidden topological angle in the    ${\cal B}_{ii} \sim e^{-2S_0 + i \pi} $ amplitude relative to the    ${\cal B}_{ij} \sim e^{-2S_0} $ amplitude \cite{Behtash:2015kna}. 
 This is the microscopic mechanism for  the cancellation of the spin-wave condensate in the   $N_f=1$  theory.

\acknowledgments
We are grateful to  Philip Argyres  for collaboration on the early stages of this work, and many fruitful discussions. 
We also thank Gokce Basar, Aleksey Cherman, Daniele Dorigoni, David Gross,  Tatsuhiro Misumi, Misha Shifman, 
Tin Sulejmanpasic, and Wojtek Zakrzewski for discussions.  MU thanks  the KITP at Santa Barbara for its hospitality during the program 
``New Methods in Nonperturbative Quantum Field Theory" where part of this work was done. 
 The authors gratefully acknowledge support from the Simons Center for
Geometry and Physics, Stony Brook University, during the Workshop on
"Resurgence and Localization in String Theory and Quantum Field Theory",
at which some of the research for this paper was performed.
GD acknowledges support from the DOE grant DE-SC0010339.
MU acknowledges support from the DOE grant  DE-SC0013036.

\end{document}